\documentclass[11pt]{article}
\usepackage{amssymb}
\usepackage{amsmath}
\usepackage{natbib}
\usepackage{graphicx}
\usepackage[linesnumbered]{algorithm2e}
\usepackage{subfigure}
\usepackage{amsthm}
\usepackage{epstopdf}
\usepackage{bbm}
\usepackage{multirow}
\usepackage{color}
\usepackage{verbatim}
\usepackage{morefloats}
\usepackage{url}

\usepackage[margin=3cm]{geometry}

\newcommand{\vect}[1]{\boldsymbol{#1}}

\begin{document}

%\date{\today}

%\title{New Insights into History Matching via Sequential Monte Carlo}
\title{A Semi-Automatic Method for History Matching using Sequential Monte Carlo}

\author{Christopher Drovandi*, David J. Nott$\dagger$ and Daniel E. Pagendam$\ddagger$ \\ \\ *School of Mathematical Sciences, \\ Australian Centre of Excellence for Mathematical and Statistical Frontiers, \\ Centre for Data Science, \\ Queensland University of Technology, Brisbane, Australia 4001  \\ email: \texttt{c.drovandi@qut.edu.au} \\ \\ $\dagger$ Department of Statistics and Applied Probability, \\ National University of Singapore \\ \\ $\ddagger$ CSIRO Data61, Dutton Park, Australia, 4102  
 }

\maketitle

\begin{center}\textbf{Abstract}\end{center}

The aim of the history matching method is to locate non-implausible regions of the parameter space of complex deterministic or stochastic models by matching model outputs with data.  It does this via a series of waves where at each wave an emulator is fitted to a small number of training samples.  An implausibility measure is defined which takes into account the closeness of simulated and observed outputs as well as emulator uncertainty.  As the waves progress, the emulator becomes more accurate so that training samples are more concentrated on promising regions of the space and poorer parts of the space are rejected with more confidence.  Whilst history matching has proved to be useful, existing implementations are not fully automated and some ad-hoc choices are made during the process, which involves user intervention and is time consuming. This occurs especially when the non-implausible region becomes small and it is difficult to sample this space uniformly to generate new training points.  In this article we develop a sequential Monte Carlo (SMC) algorithm for implementing history matching that is semi-automated.  Our novel SMC approach reveals that the history matching method yields a non-implausible region that can be multi-modal, highly irregular and very difficult to sample uniformly.  Our SMC approach offers a much more reliable sampling of the non-implausible space, which requires additional computation compared to other approaches used in the literature.

\vspace*{.3in}\noindent\textsc{Keywords}: {emulator, Gaussian process, history matching, hydrology, Markov chain Monte Carlo, Markov processes.}
\newpage

\section{Introduction} \label{sec:intro}

As practitioners strive to develop more realistic models, the computational burden of calibrating them to observed data typically increases.  The development and implementation of such models often results in some computer code that is not tractable to handle in a conventional way.  Complex models arise in many different fields such as biology (e.g.\ \citet{Vo2015}), ecology (e.g.\ \citet{Chen2017}), climate (e.g.\ \citet{Holden2015}), cosmology (e.g.\ \citet{Vernon2014}) and many other disciplines.  In the case of deterministic models, it may be very expensive to solve the model even for a single parameter configuration.  Stochastic models may be expensive to simulate, and may only produce noisy likelihood estimates.  History matching (e.g.\ \citet{Craig1997}) is a method for determining non-implausible regions of the parameter space of a complex computer model where simulation from the model is very expensive.  A non-implausible parameter value is defined to be one for which there currently is not evidence that simulation outputs generated using that value will not match observed outputs.  The method proceeds in a series of waves and uses emulation to determine thriftily without model simulation the current non-implausible region.  From this region a training sample can be formed, model simulation performed, followed by another application of emulation.  After each wave the non-implausible volume gets smaller as the emulator is more accurate over smaller volumes and more training points are generally placed closer to where simulated outputs match observed outputs.

Whilst history matching has proven to be useful in quickly eliminating large portions of the implausible parameter space,  user intervention is often required during the process and various ad-hoc decisions are made.  In this paper we develop a novel sequential Monte Carlo (SMC) algorithm for history matching.  By doing this, we obtain a more generic, semi-automated (less practitioner tuning) and principled algorithm for exploring the non-implausible space.  As a by-product of the latter, we demonstrate that the non-implausible region generated from history matching can be extremely complex and difficult to explore even for sophisticated sampling methods.  This serves to highlight challenges for practitioners that can arise from history matching and provides motivation for further research in this area.    

The paper is structured as follows.  Details of the history matching approach and some variants of how it is implemented are provided in Section \ref{sec:history_match}.  Section \ref{sec:SMC_history_match} describes the novel SMC framework for history matching.   Numerical examples to illustrate the ideas in this paper are shown in Section \ref{sec:Examples}.  The paper concludes with a discussion in Section \ref{sec:Discussion}.

\section{History Matching} \label{sec:history_match}

In this section we describe the history matching method and some of the issues with current implementations of it.  We note that history matching is sometimes referred to as a general problem rather than a method (e.g.\ \citet{Oliver2011}).  Our usage of the term `method' is consistent with some other terms in the literature such as `technique' \citep{McKinley2017} and `iterative procedure' \citep{Andrianakis2015}. 

\subsection{The Method}

Let $\theta \in \Theta \subseteq \mathbb{R}^p$ be the $p$-dimensional parameter of interest of a model $\mathcal{M}$.    An initial non-implausible region of interest $\Theta_0$ is defined.  The model $\mathcal{M}$ may be deterministic or stochastic.  We denote the solution of the deterministic system as $y_\theta = \mathcal{M}(\theta)$ and generation from the stochastic system as $y_\theta = \mathcal{M}(\theta,u)$ where $u$ are the random numbers required in the stochastic model and $y_\theta \in \mathcal{Y} \subseteq \mathbb{R}^n$ where $n$ is the number of outputs.  We write $y_\theta^k$ as the $k$th output of the model for $k=1,\ldots,n$.  We use the term `model simulation' for both deterministic and stochastic models.  The observed output is denoted as $y_{\mathrm{obs}} \in \mathcal{Y}$.

\begin{algorithm}
	\SetKwInOut{Input}{Input}
	\SetKwInOut{Output}{Output}
	\caption{Steps involved in the history matching algorithm.}
	\label{alg:history_match}
		\Input{An initial non-implausible region of interest $\Theta_0$.}
		\Output{A region $\Theta_s \subset \Theta$ that is deemed as non-implausible.} 
		Set the wave counter to $w=1$ \\
		Generate $N_w$ training samples from $\Theta_0$ using a space filling design and simulate the model at each $\theta_j$ to generate the collection of outputs $\{y_{\theta_j}\}_{j=1}^{N_w}$. \\
		Determine a set of outputs to be emulated at wave $w$, denoted $Q_w \subseteq \{1,2,\ldots,n\}$.   Fit an emulator $E_w^k$ to the training data $\{\theta_j,y_{\theta_j}^k\}_{j=1}^{N_w}$ for $k \in Q_w$. \label{line:emulate} \\
		Use the emulator $E_w^k$ to define an implausibility function $\mathcal{I}_w^k(\theta)$ for $k \in Q_w$.  If $\mathcal{I}_w^k(\theta) > c_w$ for some chosen $c_w$ for any $k \in Q_w$, then $\theta$ is deemed as implausible by emulator $E_w$. \\
		Use all emulators $E_r^k$ for $r=1,\ldots,w$ and $k \in Q_r$  to define the non-implausible region $\Theta_w = \cap_{r = 1}^w \cap_{k \in Q_w} \{\theta \in \Theta |  \mathcal{I}_r^k(\theta) < c_r  \}$. \\
		Increase wave counter $w = w+1$. \\
		Generate $N_w$ training samples $\{\theta_j\}_{j=1}^{N_w}$ from $\Theta_w$ and simulate the model at each $\theta_j$ to generate the collection of outputs $\{y_{\theta_j}\}_{j=1}^{N_w}$. \\ 
		If the stopping rule is satisfied then finish otherwise return to Line \ref{line:emulate}. \\
	
\end{algorithm} 

The steps involved in history matching are shown in Algorithm \ref{alg:history_match}.  The first step involves generating a training sample from $\pi(\theta)$ and simulating from the model at these points.   It is common in the history matching literature to use an approach with improved space filling properties compared to pseudo-random samples.  For example, latin hypercube sampling \citep{Iman2008} or quasi-Monte Carlo (QMC, \citet{Niederreiter1992}) are popular choices.  

An emulator is then fitted to the training sample for some subset of the outputs.  We denote the set of outputs emulated at wave $w$ as $Q_w \subseteq \{1,2,\ldots,n\}$.  The set $Q_w$ usually includes the outputs that can be emulated with reasonable accuracy.  Often the set $Q_w$ will grow in later waves as more of the non-implausible space is reduced, allowing some outputs to be more easily emulated \citep{Vernon2018}.    For the moment we drop the output index $k$ and consider a single output for notational convenience.  A common emulator choice is the Gaussian process (GP), but in principle any emulator can be used.  In the numerical results in Section \ref{sec:Examples} we use a GP and assume that the reader is familiar with them (see \citet{Rasmussen2006} for details on GPs).  The fitted emulator can then divide the parameter space into implausible and non-implausible regions.  The practitioner must decide upon an implausibility measure, $\mathcal{I}(\theta)$, for each output to inform the split.  In particular, an untested $\theta$ is deemed as currently non-implausible if $\mathcal{I}(\theta) < c$ for some chosen cut-off $c$.  For multiple outputs, there is an implausibility measure for each and a non-implausible test value must satisfy the cut-off for all outputs.   One common choice of the implausibility measure (e.g.\ \citet{Andrianakis2015}) is as follows
\begin{align}
\mathcal{I}(\theta) &= \frac{|y_{\theta} - y_{\mathrm{obs}}|}{\sqrt{s_{m,\theta}^2 + s_{e,\theta}^2 + s_{d}^2 + s_{\mathrm{obs}}^2} } \label{eq:implausibility},
\end{align}  
where $s_{m,\theta}$ is the estimated standard deviation of the model output, $s_{e,\theta}$ is a standard deviation that arises from emulation uncertainty, $s_{\mathrm{obs}}$ is the standard deviation of the observation error, and $s_{d}$ is an additional standard deviation to account for the fact that the model might be misspecified in some way.  The model output standard deviation $s_{m,\theta}$ is zero if the model is deterministic.  In the case of a stochastic model, the model output standard deviation can be estimated easily from the training data if it is assumed that it is independent of $\theta$.   If this is not a reasonable assumption, then $K$ independent simulations can be performed for each $\theta$ in the training sample, and a surface can be fitted to the empirical standard deviations to predict for untested $\theta$.  When $K$ simulations are performed, the output $y_{\theta}$ is an average of the $K$ simulations.  Choosing $c=3$ can be justified using Pukelsheim's rule \citep{Pukelsheim1994}.  Pukelsheim's rule is the powerful result that states that for any continuous, unimodal distribution, 95\% of its
probability must lie within $\pm 3\sigma$, regardless of asymmetry, skew, or heavy tails.  In the case of multiple outputs, \citet{Vernon2018}, for example, incorporate an additional step that, for each wave, includes only the emulators that are deemed as being accurate enough.

The implausibility measure could also be formed from the log-likelihood function of the underlying model.  In latent variable stochastic models, an estimate of the likelihood could be obtained via importance sampling methods \citep{Andrieu2009}, which may be expensive to obtain.  In such instances we could consider the implausibility function as $\mathcal{I}(\theta) = -\log f(y_{\mathrm{obs}}|\theta) - r \times s_{e,\theta}^2$ and recognise that the log-likelihood, $\log f(y_{\mathrm{obs}}|\theta)$, may not be the exact value but rather only an estimate of it.  It is natural to consider the log-likelihood as opposed to the likelihood as the former is generally easier to emulate as the curvature of the log-likelihood surface does not vary as greatly with respect to the parameter.  Further, for numerical stability, the log-likelihood may be all that is available.  \citet{Wilkinson2014} consider even taking the log a second time if the log-likelihood values differ widely for different $\theta$ in the training sample.  The value of $r$ may be considered as an exploration parameter.  The larger it is the more we are inclined to explore where the emulator is uncertain.  The smaller it is the more we exploit regions where the emulator predicts the log-likelihood to be large.  Finally, if we simply take the distance between observed and simulated data we might consider $\mathcal{I}(\theta) = \rho_\theta - r \times s_{e,\theta}^2$.  Based on the Gaussian assumption of the GP prediction we may take $r=3$. 

The choice of what type of output(s) to use in history matching is likely to be problem dependent.  The advantage of emulating a log-likelihood function is that only a single output needs to be emulated.  However, given it is a function of several model outputs combined via a
probabilistic model for the data generating process, the log-likelihood surface may be a complex and multi-modal function which is difficult to emulate and will also lead to a more challenging non-implausible region to sample.  In many applications, separate model outputs might be less complex functions of the input parameter and easier to emulate, but more emulators are required.  For more discussion on these points, see \citet{Vernon2010, Vernon2010Rejoinder, Vernon2018}.    The choice of model output is not the focus of this paper.  In Section \ref{sec:Examples} we consider examples involving a single model output (a distance or log-likelihood function) and an example with several model outputs.  Instead, our motivation is to demonstrate non-implausible regions that are difficult to sample from and are thus suitable test examples for our new method.

The next step involves generating uniformly from the non-implausible parameter space, which we denote as $\Theta_1 = \cap_{k = 1}^n \{\theta \in \Theta | \mathcal{I}^k(\theta) < c_1 \}$, re-introducing again the possibility of $n>1$ outputs.  This produces a new training sample that is concentrated in more promising parts of the parameter space relative to the initial training sample.  This essentially re-starts the history matching process, where an emulator is fitted to each output of the new training sample.  The new emulators are likely to lead to more accurate predictions as the training sample is less diffuse and the outputs simulated from the model less variable across $\Theta_1$.  This often results in a smoother function and reduced model stochasticity, enabling more accurate emulators.  The history matching method continues in this fashion in waves, where we denote the wave counter as $w$.  At wave $w$ the implausibility function for output $k$ is denoted $\mathcal{I}_w^k(\theta)$ with a cut-off $c_w$ potentially depending on the wave.  After $w$ waves, the non-implausible parameter space is denoted as $\Theta_w = \cap_{r = 1}^w \cap_{k = 1}^n \{\theta \in \Theta |  \mathcal{I}_r^k(\theta) < c_r  \}$.  That is, for an untested $\theta$ to be deemed as non-implausible for wave $w+1$ it must be deemed as non-implausible for all previous $w$ waves.   

The waves continue until a user-specified stopping rule is met.  \citet{Andrianakis2017} discuss some commonly used stopping rules; the entire space may be deemed implausible, the emulator's uncertainty is small enough and/or a sufficient number of points that match the observed data have been collated.  Further, when the implausibility measure is based on a distance or approximate likelihood function  there is no easily defined cut-off of implausibility nor stopping rule, unlike the standard
history matching implausibility measures given by \eqref{eq:implausibility} and its variants.  To simplify the exposition, we do not attempt to address the issue in this paper, except that we can stop the algorithm if it becomes too computationally demanding to sample uniformly from the non-implausible region.  Further, given the semi-automatic nature of our algorithm, it is possible to leave the algorithm running and save the output after each wave, and then decide on a stopping rule post-hoc.

For simplicity in this paper we assume there is either a single output $n=1$ or that the implausibility measures of $n>1$ outputs are combined into a single implausibility measure, such as the second maximum implausibility measure \citep{Vernon2010}.  We find that such instances are sufficient to demonstrate the performance of the algorithm and also the challenges that can arise from history matching.  The single output could be some scalar $y_{\mathrm{obs}}$.  Alternatively, when $y_{\mathrm{obs}}$ is vector-valued, the output could be some distance $\rho_\theta = ||y_\theta-y_{\mathrm{obs}}||$ between simulated and observed outputs, or the `distance' could be quantified through a chosen likelihood function, $f(y_{\mathrm{obs}}|\theta)$ or a 1-1 transformation of it such as the log-likelihood.  If the exact likelihood computation is not feasible, an approximate likelihood could be used, such as a stochastic estimator of the likelihood.  In Section \ref{sec:Discussion} we discuss how our SMC algorithm for history matching could be extended to multiple outputs, whose implausibility measures are not combined into a single implausibility measure.

\subsection{Issues with History Matching}

Despite the ability of history matching to relatively quickly identify parts of the parameter space that may be consistent with the observed data, it does have at least two issues:
\begin{enumerate}
	\item The cut-off values $c_w$ may not be easy to select in practice and there is no existing automated method for doing so.
	\item Sampling uniformly from $\Theta_w$ as $w$ increases becomes increasingly difficult.
\end{enumerate}

Regarding issue 1, as was already mentioned, a sensible cut-off value for the implausibility measure in \eqref{eq:implausibility} is 3.  However, this value assumes that we can accurately compute the quantity in \eqref{eq:implausibility}.  This would involve being able to model $s_{m,\theta}$ and $s_{d}$ accurately, which may be difficult to do, especially for the model misspecification term $s_{d}$ (especially when this term may depend on $\theta$).  Sometimes the $s_d$ term is simply ignored.  If the implausibility measure is not calculated accurately enough then a cut-off value of $c=3$ may lead to the non-implausible parameter space reducing too quickly or not quickly enough.  A conservative estimate of $s_d$ can be used to make the choice $c=3$ meaningful \citep{Andrianakis2015}, but it may not be clear what degree of conservatism is appropriate.  Many authors (e.g.\ \citet{Vernon2014}, \citet{Wilkinson2014} and \citet{Andrianakis2015}) report instances of having to manually change these cut-off values after each wave and/or resort to differing numbers of training samples for each wave.  Regarding the log-likelihood implausibility measure, \citet{Wilkinson2014} suggest a cut-off value that depends on the maximum log-likelihood value/estimate in the training sample minus 10 log-likelihood points, but this value may not be appropriate for all applications and may also need to change over the waves.  Moreover, an approximate cut-off based on the log-likelihood is likely to depend on the parameter dimension, and also potentially on the number of training samples, since a larger number of training samples will increase the chance of landing near the maximum log-likelihood.  If the implausibility measure is based on some distance between observed and simulated data, then an appropriate cut-off value is generally unclear and will need to change throughout the algorithm as distances get generally smaller.  It could be argued that an implausibility metric such as \eqref{eq:implausibility} should be strongly preferred precisely because this gives the cut-off a meaningful interpretation, but there will be times when a one-dimensional distance measure is practically convenient.  Thus an approach able to select the thresholds in a more automated way would be welcome. 

Regarding issue 2, uniform samples from $\Theta_w$ can be obtained perfectly by rejection sampling; continually sampling from $\pi(\theta)$ until $\theta \in \Theta_w$ (see, for example, \citet{Wilkinson2014}).  However, as $w$ increases, the acceptance rate of this rejection sampler decreases rapidly.  It may become unacceptably small if the volume of the parameter space consistent with the observed data is a tiny fraction of the volume of the original parameter space.  Once the rejection sampler becomes too inefficient various authors resort to ad-hoc approaches for generating training samples from subsequent waves, choices that do not preserve the uniform distribution on $\Theta_w$.  For example, \citet{Andrianakis2015} proposes 20 parameter values centered on a subset of parameter values from a previous wave that are deemed non-implausible.  An approach that can reliably sample the non-implausible space uniformly is of interest to ensure that the parameter space is comprehensively explored.  

\citet{Williamson2013} consider a more sophisticated approach to the sampling problem based on evolutionary Monte Carlo.  The approach is similar to parallel tempering in that several chains are run in parallel.  One of the chains has the desired target, the uniform distribution over $\Theta_w$, as its limiting distribution, whilst the other chains increasingly relax the constraints so that they can more freely search the parameter space.  The method has local moves to update each chain and also proposes swaps between chains.  Although the approach has the desired target as its limiting distribution, its implementation is not straightforward and there are many tuning parameters such as the number of chains, the target distribution of each chain, local proposal distributions and chain swapping mechanisms.  

\citet{Andrianakis2017} provide another serious attempt to uniformly sample the non-implausible region.  The method is based on slice sampling.   However, since each co-ordinate of the input is updated separately, this approach may be inefficient if there is dependence between inputs implied by the data.  Nonetheless, the approach of \citet{Andrianakis2017}, or some adaptation of it, could be used in our SMC framework described next.

\section{SMC Framework for History Matching} \label{sec:SMC_history_match}

Here we propose to place the history matching method into the SMC framework.  This allows us to take advantage of the extensive research on efficient SMC algorithms (e.g.\ \citet{DelMoral2006}, \citet{Fearnhead2013} and \citet{South2016}), which are naturally adaptive.   

SMC involves sampling from a sequence of distributions that smoothly evolves between a distribution that is easy to sample from and finishing at the distribution of interest, sometimes referred to as the target distribution.  For history matching, we define the sequence of distributions as
\begin{align}
p_w(\theta) \propto \pi(\theta)\prod_{k=1}^w\mathbb{I}(\mathcal{I}_k(\theta) \leq c_k), \label{eq:sequence}
\end{align}
where $\pi(\theta)$ defines the distribution that samples are initially drawn from, for example a uniform distribution over a hyper-rectangle or some other distribution that has been informed from experts or historical data.  As was mentioned previously, history matching usually draws the first set of training samples from an initial region of interest using a space filling design.  We note this is not equivalent to drawing pseudo-random numbers from a uniform distribution.  If desired, we can initialise the SMC process in the same way as standard history matching, in which case $\pi(\theta)$ is not the density of a statistical distribution but we retain it for notational convenience.   Our approach fits within the framework of \citet{Chopin2002}, but see \citet{DelMoral2006} for a more general framework for SMC methods.  Assume that we have a collection of weighted samples or `particles', $\{W_w^i,\theta_w^i\}_{i=1}^M$ from $p_w(\theta)$.  To push the particle set to the next target, a re-weighting step is required.  A simple importance sampling argument leads to the following update of the weights
\begin{align*}
W_{w+1}^i \propto W_{w}^i  \mathbb{I}(\mathcal{I}_{w+1}(\theta_w^i) \leq c_{w+1}),
\end{align*}
so that the values of the samples $\theta_w^i$ themselves remain unchanged.  Assuming that $W_{w}^i = 1/M$ for $i = 1,\ldots,M$, the weights $W_{w+1}^i$ will either be proportional to a constant or equal to zero.  Thus placing the history matching approach into the SMC framework allows us to select the implausibility cut-off at each wave $c_{w+1}$ so that a certain proportion, $\alpha$, have a non-zero weight.  This is equivalent to ensuring that the effective sample size (ESS), often measured by $1/\sum_{i=1}^M (W_{w+1}^i)^2$, is roughly $\alpha M$.  A similar approach is adopted by \citet{DrovandiBiom2011} and \citet{DelMoral2008} in the SMC algorithms developed for approximate Bayesian computation.  \citet{Drovandi2016} develop a similar idea for calibrating differential equations to population data in the presence of parameter uncertainty.

After the re-weighting step the ESS drops to roughly $\alpha M$.  Resampling $M$ times from the surviving particles reproduces an equally weighted sample of size $M$.  Although mathematically the ESS is equal to $M$ after resampling, the drawback is that some particles will be duplicated.  A diverse sample from each of the targets is desired.  This can be achieved by applying an MCMC kernel to each of the resampled particles.  Given that an MCMC kernel may reject proposals, it is generally advised to apply say $R$ iterations of the MCMC kernel.  We implement the adaptive strategy promoted by \citet{South2016}, which involves running one MCMC iteration on each particle, estimating the average MCMC acceptance rate, and using that to determine $R$.

The selection of an efficient MCMC kernel is critical to the success of the SMC algorithm.  SMC has the distinct advantage over serial-based algorithms such as standard MCMC in that the population of particles can be harnessed to help build a useful MCMC proposal.  The simplest approach is a multivariate normal random walk (perhaps after transforming each parameter to the real line) with a covariance matrix calibrated from the weighted or re-sampled particle set (e.g.\ \citet{DrovandiBiom2011}).  However, given the complexity of the distributions that arise from history matching (illustrated in Section \ref{sec:Examples}), we find that this standard approach is infeasible due to the low MCMC acceptance rates generated and hence unattractively large values of $R$.

The slice sampling approach of \citet{Andrianakis2017} could be used in the MCMC step of our method.  However, we find some success for the examples in this paper with the following approach.  A kernel density estimate (kde) using the Epanechnikov kernel is calibrated to each component of $\theta$ (i.e.\ the marginals) based on a subset of the $M$ re-sampled particles.  If the parameter space is bounded then the kde is also restricted to that support.  The cumulative distribution function of the fitted kdes can be used to transform each of the marginals to roughly standard uniform, which can be transformed again to roughly standard normal via the standard normal quantile function.  A multivariate normal random walk is then applied to this transformed space.  The Jacobian term of the transformation is also incorporated into the Metropolis-Hastings ratio.  We find that the distribution on the transformed space is significantly more regular compared to the transformed space, in the sense that heavy tails can be reduced and modes brought closer together.  Once a proposal is generated on the transformed space, a sample on the original space requires evaluating the quantile function for each of the fitted marginals.  Unfortunately the quantile function of the kde has no explicit form, but it can be approximated numerically.  All of these options and functionality are provided in the \verb|ksdensity| function in Matlab.

We note that the requirement to approximate the quantile function of the kde for each of the marginals substantially slows down the MCMC step relative to other standard transforms such as the log (positive parameters) or the logistic (bounded parameters).  However, we find that in the challenging examples in Section \ref{sec:Examples} the approach we adopt is critical to maintain a reasonable value of $R$ throughout the algorithm.  More detail and comparison is provided in Section \ref{sec:Examples}.  In typical applications of history matching where the simulation of the model is very expensive, the MCMC step of the SMC algorithm, which does not involve any model simulation, may remain relatively fast.  However, one of the messages of this paper is that a significant and perhaps unavoidable amount of effort needs to be spent on ensuring the non-implausible part of the space is well represented at each wave.

Because the sequence of distributions \eqref{eq:sequence} implied by SMC history matching involves indicator functions, it is possible to reject MCMC proposals early based on $\pi(\theta)$ and proposal densities before checking for implausibility from the fitted emulators.  This idea has also been used in the context of ABC \citep{Picchini2014}.  Furthermore, once a proposal has been deemed implausible by one of the emulators, there is no need to check for implausibility with the remaining emulators.

After each wave, the SMC process generates $M$ particles from $p_w(\theta)$.  The training sample for the next wave can be obtained by sub-sampling $N << M$ particles from this set (with duplicates ignored for deterministic models).  The SMC history matching algorithm developed in this paper is summarised in Algorithm \ref{alg:SMC}.  

The larger the value of $M$ the more chance there is of the non-implausible space being well represented.  On the other hand, the computing time of the MCMC step will increase linearly with $M$.

It would be possible to implement the algorithm with \emph{a priori} fixed implausibility thresholds.  However, the proportion of the SMC particles satisfying the threshold will be unknown, and could be very small or 0.  It is important that there are a sufficient number of `alive' particles to inform the proposal distribution for the MCMC step.  In the examples below we often take $M=10,000$ and $\alpha = 0.5$ for illustrative purposes, which are rather conservative choices.  Selecting $\alpha = 0.5$ does not aim to rapidly eliminate regions of the input space, and thus is more suitable for computer models that are relatively fast.  For more expensive computer models, it will be of interest to choose a smaller $\alpha$ to more aggressively eliminate implausible regions and reduce the number of expensive runs needed.  It is important though to ensure there are enough `alive' particles at each wave to inform the MCMC proposal.  One way to counteract a smaller $\alpha$ is to increase $M$.  An alternative approach would be to run additional SMC iterations to more gradually target the non-implausible region implied by the current SMC history matching wave.  These additional SMC iterations would not trigger the expensive computer model.

 \begin{algorithm}[H]
 	\SetKwInOut{Input}{Input}
 	\SetKwInOut{Output}{Output}
 	\Input{The number of SMC particles $M$, the initial distribution $\pi(\theta)$,  the desired probability $1-c$ of moving a particle in the MCMC step, the proportion $\alpha$ of SMC particles to keep at each iteration that also helps to define the implausibility cut-off and the number of training samples $N$ for fitting the emulator at each wave.}
 	\Output{A collection of parameter values $\{\theta^i\}_{i=1}^M$ that are deemed as non-implausible.}
 	Set wave counter $w = 0$\\
 	Set $W_{0}^{i}=\frac{1}{M}$ for $i=1,\ldots,M$\\
 	Simulate $\theta_{0}^{i} \stackrel{\mathrm{iid}}{\sim} \pi(\theta)$ for $i=1,\ldots,M$ \\
 	Simulate $N$ training samples from $\pi(\theta)$ using a space filling design and fit an emulator $E_w$\\
 	\While{stopping rule not met}{
 		Set $w = w+1$\\
 		Compute implausibility measure $\mathcal{I}_w(\theta_{w-1}^{i})$ for $i = 1,\ldots,M$ \\
 		Determine implausibility cut-off value $c_w$ based on the $\alpha$ quantile of implausibility values $\{\mathcal{I}_w(\theta_{w-1}^{i})\}_{i=1}^M$\\
 		Resample $\text{floor}((1-\alpha) M)$ particle values from the surviving particles (i.e.\ those with $\mathcal{I}_w(\theta_{w-1}^{i}) \leq c_w$) to replenish the SMC population\\
 		Set $\theta_{w}^{i} = \theta_{w-1}^{i}$ for $i = 1,\ldots,M$\\
 		Form a suitable MCMC proposal distribution from the resampled particles\\
 		Move $\{\theta_w^i\}_{i=1}^{M}$ with one iteration of an MCMC kernel targetting $p_w(\theta)$ (see Algorithm \ref{alg:move})\\
 		Compute $R_t=\left\lceil \frac{\log{(c)}}{\log{(1-p_{\mathrm{acc}})}}\right\rceil$ where $p_{\mathrm{acc}}$ is the acceptance probability of the above move step\\
 		\For{$k = 1$ to $R_t$}{
 			Move $\{\theta_w^i\}_{i=1}^{M}$ with one iteration of an MCMC kernel targetting $p_w(\theta)$ (see Algorithm \ref{alg:move})\\
 		}
 		Sample $N$ particles from $\{\theta_w^i\}_{i=1}^{M}$ without replacement (removing duplicates for deterministic models).  Fit emulator $E_w$ based on this training sample \\
 	}
 	\caption{SMC History Matching Algorithm.}
 	\label{alg:SMC}
 \end{algorithm}

\newpage
 
 \begin{algorithm}[H]
 	\SetKwInOut{Input}{Input}
 	\SetKwInOut{Output}{Output}
 	\Input{Collection of re-sampled particles $\{\theta^i\}_{i=1}^M$ from $p_w(\theta)$, wave counter $w$, fitted emulators $\{E_1,E_2,\ldots,E_w\}$, the implausibility threshold cut-offs $c_1,c_2,\ldots,c_w$ and the initial distribution $\pi(\theta)$.  For simplicity of notation we omit the wave index $w$ from $\theta$.}
 	\Output{Collection of particles $\{\theta^i\}_{i=1}^M$ from $p_w(\theta)$ that are better diversified.}
 	Fit a kde to each marginal of $\theta$ based on the collection of particles $\{\theta^i\}_{i=1}^M$ \\
 	Use the cdf of the fitted kde and the normal quantile function to transform each margin to roughly standard normal.  Denote the transformed particles as $\{z_w^i\}_{i=1}^M$ \\
 	Estimate covariance matrix from $\{z^i\}_{i=1}^M$, $\hat{\Sigma}$ \\
 	\For{$i = 1,\ldots,M$}{ \label{line:forloop}
 		Propose $z^* \sim \mathcal{N}(z^i, \hat{\Sigma})$ \\
 		Use the fitted kdes to transform each margin of $z^*$ back to the original scale denoted $\theta^*$\\
 		Compute the proposal density $q(\theta^*|\theta^i) = \exp \left( \sum_{k=1}^p \log \hat{f}_k(\theta^*[k]) - \sum_{k=1}^p \log \mathcal{N}(z^*[k];0,1) \right)$ where $\theta[k]$ denotes the $k$th component of $\theta$ and $\hat{f}_k(\cdot)$ denotes the fitted kde for the $k$th marginal\\
 		Compute the proposal density in the other direction $q(\theta^i|\theta^*)$\\
 		Compute the first part of the Metropolis-Hastings ratio $r = \min (\pi(\theta^*)q(\theta^i|\theta^*) / \pi(\theta^i)q(\theta^*|\theta^i))$\\
 		\If{$\mathcal{U}(0,1) > r$}{reject early and go to the next iteration of the for loop at line \ref{line:forloop}}
 		\For{$j = 1,\ldots,w$}{
 			Compute $\mathcal{I}_j(\theta^*)$ from fitted emulator $E_j$\\
 			\If{$\mathcal{I}_j(\theta^*) > c_j$}{
 			 	reject early and go to the next iteration of the for loop at line \ref{line:forloop}\\
 			}
 		}
 		Accept $\theta^i = \theta^*$\\
 	}
 	\caption{MCMC kernel used within the generic SMC history matching method in Algorithm \ref{alg:SMC}.  The proposal distribution for the MCMC algorithm is specific to this paper.  For simplicity of notation we omit the wave index $w$ from $\theta$.}
 	\label{alg:move}
 \end{algorithm}
 
One advantage of the SMC sampling approach over the evolutionary Monte Carlo method of \citet{Williamson2013} is that the information from the SMC population of particles at the current wave can be harnessed to help facilitate uniform sampling of the non-implausible region at the next wave.  Further, the SMC approach allows for adaptive choice of the cut-offs.    Finally, the approach of \citet{Williamson2013} is an MCMC algorithm and is thus less suited to parallel computing compared to our SMC approach.  Parallel computing architectures are critical in history matching applications so that model simulations can be performed in parallel for different inputs.  Therefore it is useful that our SMC algorithm for uniform sampling of the non-implausible space can take advantage of parallel computing.

%\section{Comparison with Bayesian SMC Inference} \label{sec:compare}

\section{Examples} \label{sec:Examples}

Below we illustrate our ideas on a toy problem, and  non-trivial applications in hydrology, biology and reservoir modelling.  All of the substantive applications do not necessarily require history matching as the model simulation is not expensive relative to typical applications of history matching.  However, they are sufficient for us to illustrate the messages of this paper.  The first substantive example is in hydrology, and involves a deterministic model and the implausibility measure is based on a distance between observed and simulated data.  The SMC history matching approach is compared with an SMC optimisation routine.  The biological example involves a stochastic model for an autoregulatory gene network, and the implausibility measure is based on an unbiased likelihood estimator.  In this case we compare the output and efficiency with a more standard Bayesian SMC method.  The third example is in reservoir modelling, where we use the implausibility measure in \eqref{eq:implausibility}.  Again, we compare results with an SMC optimisation routine.  

For all of the examples we emulate each output with a GP with zero mean function and a squared exponential covariance function with automatic relevance determination.  The GP hyperparameters are estimated by maximum marginal likelihood estimation.  For all examples except the toy example in the next subsection we emulate the centred output using the sample mean of the expensive simulations at each wave.  This helps to improve the zero mean GP assumption.

\subsection{Toy Problem}

\begin{figure}[!htp]
	\centering
	\includegraphics[height=0.3\textheight,width=0.5\textwidth]{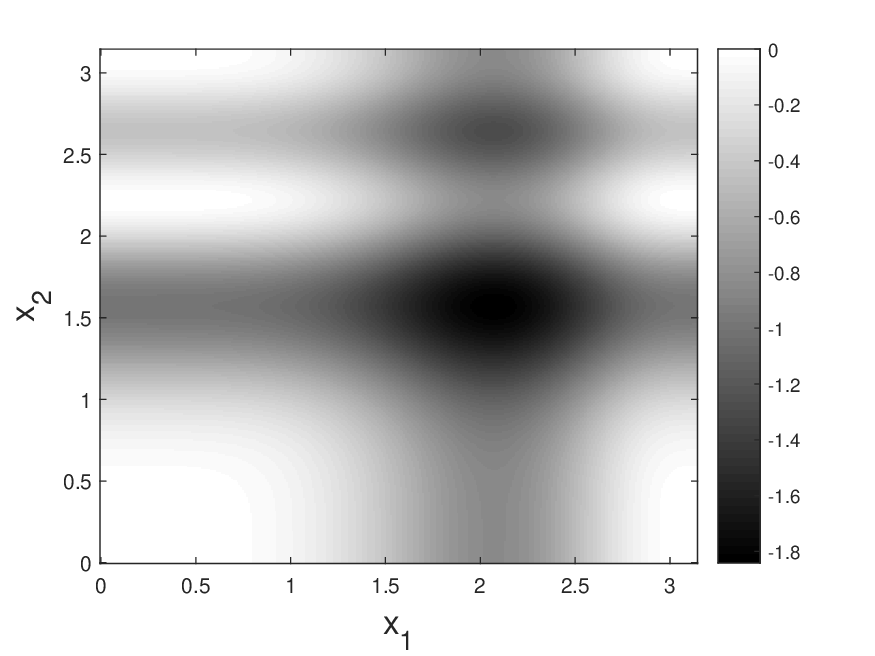}
	\caption{Visualisation of the toy function.  Black regions show where the function value is smallest.}
	\label{fig:toy_fun}
\end{figure}

Here the objective is to minimise the following function taken from one of the test functions in \citet{Molga2005}:
\begin{align*}
y &= -\sin(x_1)\sin(x_1^2/\pi)^2 - \sin(x_2)\sin(2x_2^2/\pi)^2,
\end{align*}
where $\theta = (x_1,x_2) \in (0,\pi)\times(0,\pi)$.  Figure \ref{fig:toy_fun} provides a visualisation of the function.

\begin{figure}[!htp]
	\centering
	\includegraphics[height=0.6\textheight,width=0.9\textwidth]{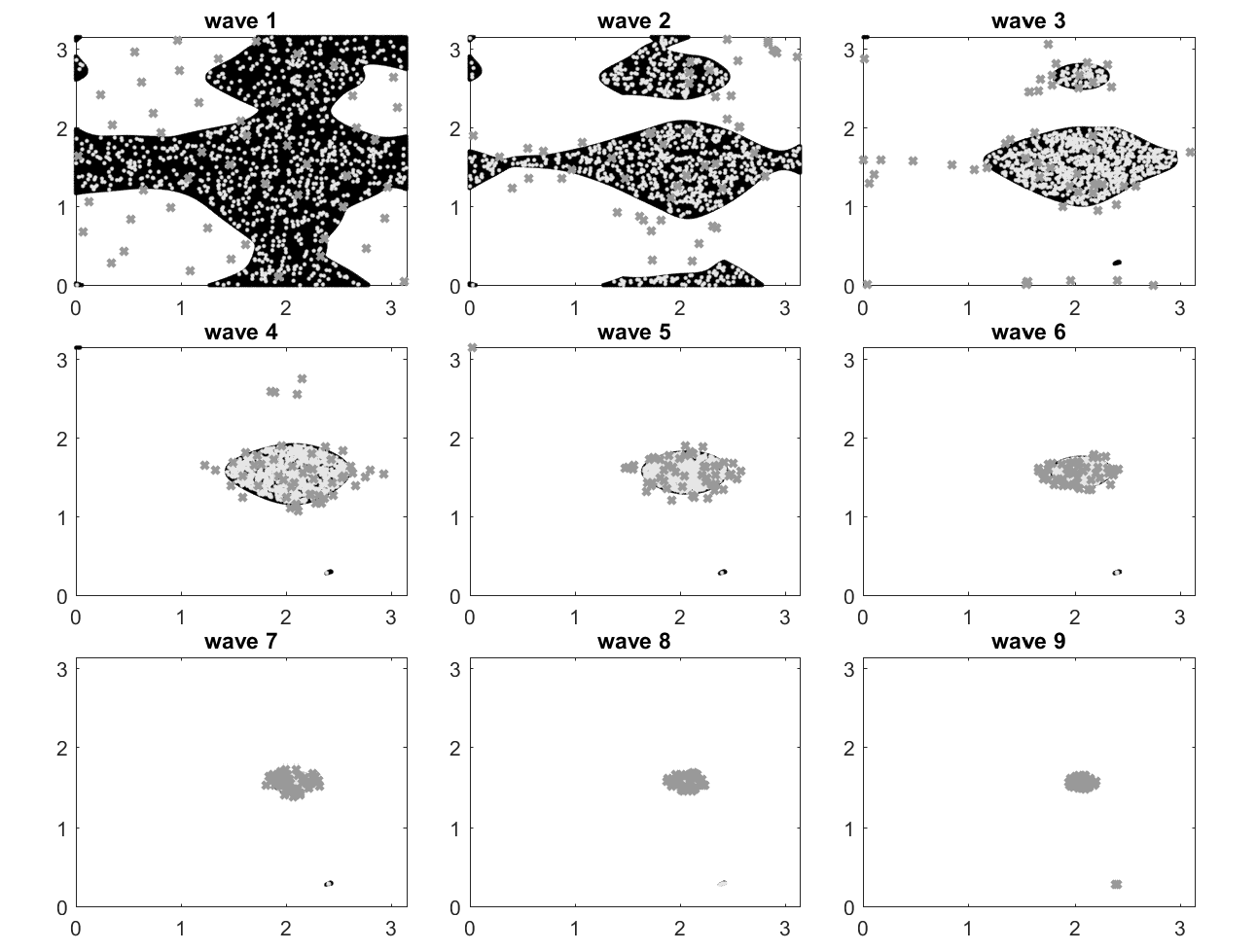}
	\caption{Results for the SMC approach to sample from the sequence of distributions obtained by brute force for the toy example.  The black regions denote the non-implausible regions after each wave and the light grey dots are the SMC samples.  The grey crosses are the training samples for fitting the GP during the brute force process.}
	\label{fig:toy_smc_bivariate}
\end{figure}

For each wave we randomly select $N=50$ training samples for fitting the GP.  The history matching method is solved using brute force by starting with $2^{20}$ QMC samples over the bounded support.  The process starts by selecting $N=50$ samples at random from this initial set and fitting a GP.  The implausibility for each sample is computed as $\mathcal{I}(\theta) =  y_p(\theta) - r \times s_p(\theta)$ where $y_p(\theta)$ and $s_p(\theta)$ is the mean prediction of the function and the standard deviation from the currently fitted GP, respectively.  The exploration parameter is set at $r=3$.  The cut-off for implausibility at wave $w$ is given by $\mathcal{I}_w(\theta) > c_w$ where $c_w$ is selected adaptively such that exactly half of the surviving samples satisfy the cut-off.  Thus after each wave half of the original $2^{20}$ samples is lost.  Nine waves are used so that there are still a significant number of the original samples that satisfy all of the waves.  In effect we have perfectly uniform samples from each of the non-implausible regions defined after each wave.  The non-implausible regions are shown in black in Figure \ref{fig:toy_smc_bivariate}.  The training samples are shown as crosses, and are effectively taken uniformly at random from the black region in the previous wave (the black region for wave 0 is the entire space of $\theta$).

What is immediately noticeable even from this very simple example is that the non-implausible region defined from the history matching procedure can be significantly disconnected, which creates a probability distribution that can be highly multi-modal and irregular, making it difficult to sample efficiently and reliably.  This is because the non-implausible regions exist where the emulator predicts low function values and/or where there is high uncertainty in the prediction.  There is a small non-implausible island at the top left of the space which exists even after wave 4, where the true function is relatively high and thus not of interest.  This part of the space can only be discarded once a training sample is placed near that region, which can be seen in wave 5.

We now fix the sequence of distributions implied by the GP fits based on perfect uniform sampling and explore the ability of SMC for sampling this sequence.  Here $M=5$K particles are used.  A random subset of $2.5$K of these samples for each wave are shown as light grey dots in Figure \ref{fig:toy_smc_bivariate}.  It is clear that the SMC procedure is quite successful at uniformly sampling the black regions.  The acceptance rate of the MCMC step remains reasonable throughout the algorithm, between 40-60\%.  In contrast, by wave 9, the acceptance rate of the brute force approach (i.e.\ perfect sampler) is 0.2\%.

\begin{figure}[!htp]
	\centering
	\includegraphics[height=0.6\textheight,width=0.9\textwidth]{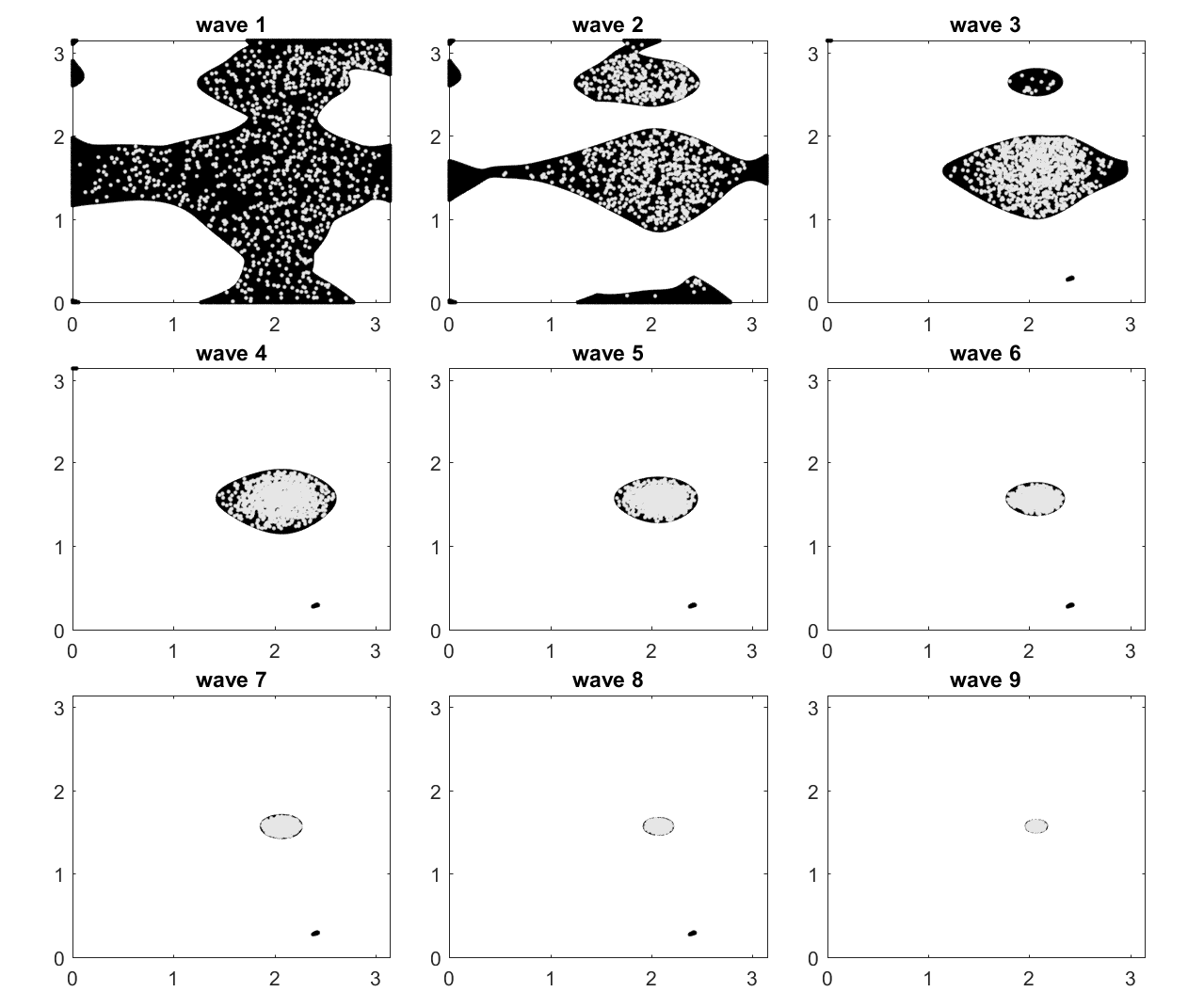}
	\caption{Results for the adhoc approach with the logit transformation to sample from the sequence of distributions obtained by brute force for the toy example.  The black regions denote the non-implausible regions after each wave and the light grey dots are samples generated by a more standard approach that might be used in history matching. }
	\label{fig:toy_adhoc_logit_bivariate}
\end{figure}

\begin{figure}[!htp]
	\centering
	\includegraphics[height=0.6\textheight,width=0.9\textwidth]{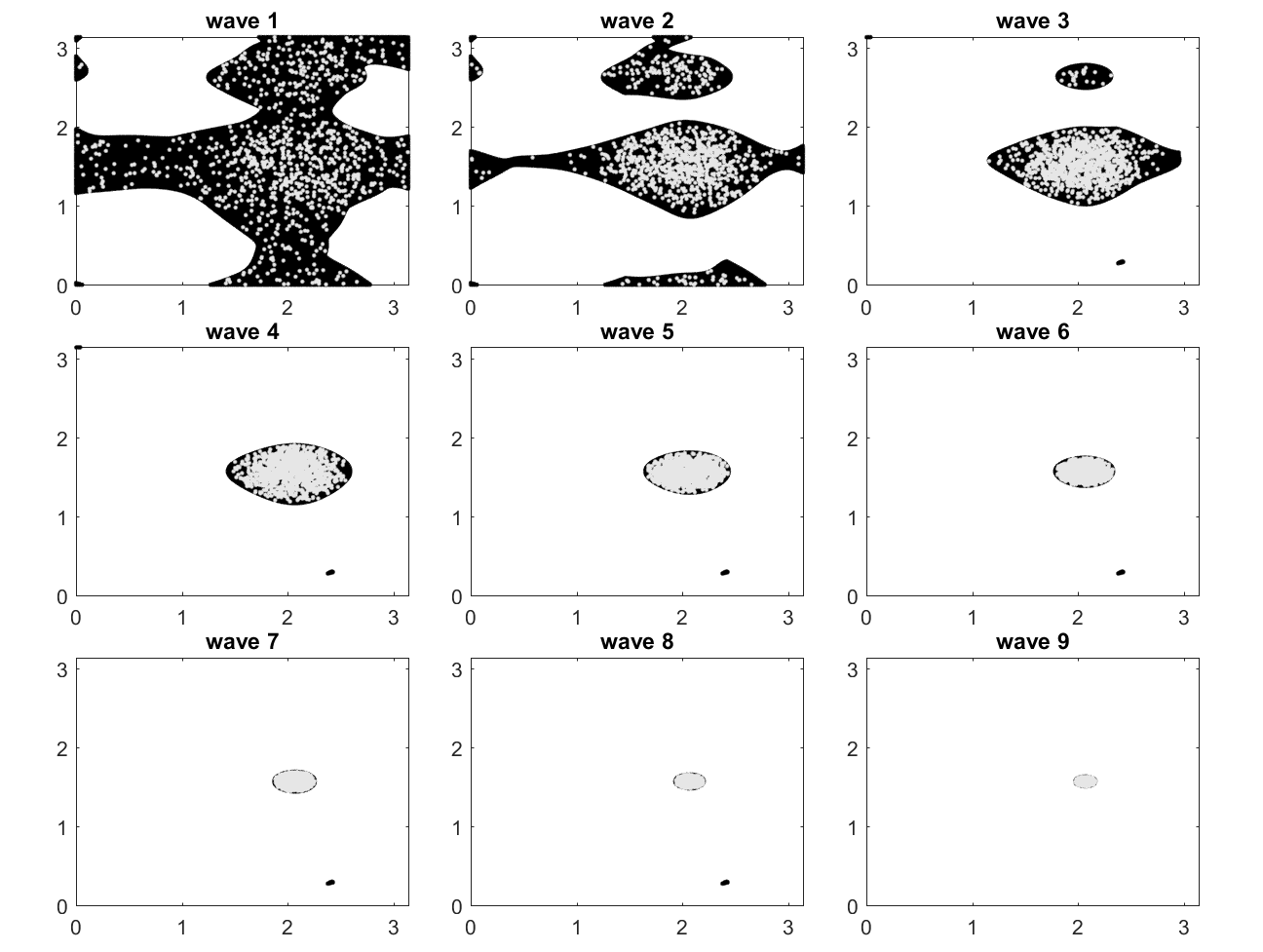}
	\caption{Results for the adhoc approach with the kde transformation applied to the marginals to sample from the sequence of distributions obtained by brute force for the toy example.  The black regions denote the non-implausible regions after each wave and the light grey dots are generated with the described ad-hoc approach. }
	\label{fig:toy_adhoc_kernel_bivariate}
\end{figure}

We compare the output of the SMC approach with a more standard approach that might be adopted in the history matching literature.  A logit transformation is applied to each component of $\theta$ to make the parameter space unbounded.  Then, the sampling distribution is specified as a multivariate normal distribution with a mean and covariance that is calculated from the existing $M=5$K particles satisfying the constraint for the next wave.  Sampling from this normal distribution continues until $N = 5$K samples are generated that satisfy all waves to date.  The samples from this approach after each wave are shown in light grey dots in Figure \ref{fig:toy_adhoc_logit_bivariate}.  It is evident from these plots that this approach under-represents and over-represents in different spots within the black region.  Some of the black regions are essentially ignored, which implies  that these regions may remain completely unexplored by the history matching algorithm, which could potentially be useful regions of the parameter space.  We also use some ideas from this paper and apply the marginal cdf transform (rather than logit transform) and fit a multivariate normal distribution on this space.  The results are shown in Figure \ref{fig:toy_adhoc_kernel_bivariate}.  The results are better than for the logit transform, but ultimately it remains clear that uniform sampling is not achieved.  The acceptance rate of the SMC approach is typically lower than the adhoc approaches, since an MCMC kernel may reject a proposal even if it satisfies all relevant waves.  However, this is a necessary price to pay to guarantee more reliable sampling of the non-implausible space.

The example here in two dimensions already demonstrates the multi-modal and irregular sequence of distributions that can be generated from the history matching process, and one can imagine how much more complex this sequence might be in higher dimensions.  The next example, which involves seven parameters, provides some insight into this.

\subsection{Hydrology Example}

Here the objective is to find parameter values of a rainfall-runoff model (RRM) that lead to good predictions of a stream flow time-series given input time series of rainfall and potential evapotranspiration.  RRMs are hydrological models that conceptualise the process by which precipitation across a catchment is transformed into water flowing in a river or stream (surface water).  The mechanics of an RRM are typically described by a set of water balance equations (differential or difference equations), involving a number of conceptual reservoirs or stores of water in a catchment.  In this example, we consider a simple, spatially-lumped (in the sense that the conceptual water stores and precipitation are not modelled as spatially-distributed processes) RRM used by \cite{Schoups2010a} and \cite{Schoups2010b}, which is derived from the FLEX model of \cite{Fenicia2007}.

The RRM consists of four conceptual reservoirs of water in a catchment: (i) an interception reservoir that accounts for precipitation intercepted by vegetation; (ii) a soil water reservoir (the unsaturated zone); (iii) a fast-reacting reservoir with relatively short residence time, so that water in this store appears as surface water relatively quickly; and (iv) a slow-reacting reservoir, for which there is a relatively long residence time before water in this store emerges as surface water.  The stores identified in (iii) and (iv) give rise to surface water components that are sometimes referred to by hydrologists as quickflow and slowflow (or baseflow).  

The states of the four reservoirs (interception, unsaturated, fast and slow) at time $t$ are denoted $I_t, U_t, F_t$ and $S_t$ respectively, and in order to remain independent of the catchment area, these storages are measured in the same units as rainfall (mm).  The interception reservoir intercepts precipitation (measured) which enters at rate $P_t$ (mm/day), and can fill to a maximum storage capacity of $I_{\mathrm{max}}$.  Because some of the precipitation over the catchment is retained in the interception reservoir, the effective precipitation rate, $P^{_e}_{t}$ (mm/day), entering the soil water store, is less than the measured precipitation and is calculated as $P^{_e}_{t} = P_t - (I_{\mathrm{max}} - I_t)$.  The interception reservoir also loses water to the atmosphere via evapotranspiration at a rate that is calculated as $E^{_I}_{t} = \min(E^{_p}_{t}, I_t)$ (mm/day), where $E^{_p}_{t}$ is the potential evapotranspiration rate (mm/day), typically calculated based on environmental factors (temperature, solar radiation, wind speed etc) using the Penman-Monteith equation \citep{Monteith1965}.  The water balance for the interception reservoir is therefore
\begin{align*}
\frac{dI_t}{dt} &=  P_t - E^{_I}_{t} - P^{_e}_{t}.
\end{align*}
Our model assumes that the soil water store, $U_t$, has a maximum capacity of $U_{\mathrm{max}}$ (mm), and its water balance is governed by the equation
\begin{align*}
\frac{dU_t}{dt} = P^{_e}_{t} - Q^{_f}_{t} - E^{_a}_{t} - Q^{_s}_{t},
\end{align*}
where $P^{_e}_{t}$ is the effective rainfall rate at time $t$ as previously defined, $Q^{_f}_{t}$ is the runoff rate (mm/day) that enters the fast-reacting reservoir, $E^{_a}_{t}$ is the actual evapotranspiration (mm/day) and $Q^{_s}_{t}$ (mm/day) is the percolation rate (mm/day) into the slow-reacting reservoir.  The fluxes $Q^{_f}_{t}, Q^{_s}_{t}$ and $E^{_a}_{t}$ are modelled as
\begin{align*}
Q^{_f}_{t} &= P^{_e}_t f\left(\tfrac{U_t}{U_{\mathrm{max}}}; \alpha_F\right), \\
E^{_a}_{t} &= (E^{_p}_t - E^{_I}_t) f\left(\tfrac{U_t}{U_{\mathrm{max}}}; \alpha_E\right), \quad \textup{and} \\
Q^{_s}_{t} &= Q^{_s}_{\mathrm{max}}f(\tfrac{U_t}{U_{\mathrm{max}}}; \alpha_S),
\end{align*}
where $f(U; \alpha) = \frac{1 - e^{-\alpha U}}{1 - e^{- \alpha}}$ is a sigmoidal function that is monotonically increasing in $U$.  $Q^{_s}_{\mathrm{max}}$ is the parameter defining the maximum percolation rate into the slow-reacting reservoir, and $\alpha_F, \alpha_E$ and $\alpha_S$ ($\neq 0$) are parameters governing the rate of change of the sigmoidal function for each of the three fluxes.  The fast-reacting reservoir and slow reacting reservoirs have water balances, respectively of
\begin{align*}
\frac{dF_t}{dt} &= Q^{_f}_{t} - Q^{_F}_{t} \quad \textup{and} \\
\frac{dS_t}{dt} &= Q^{_s}_{t} - Q^{_S}_{t},
\end{align*}
where $Q^{_F}_{t} = K_F F_t$ (mm/day) and $Q^{_S}_t = K_S S_t$ (mm/day).  Finally, streamflow is then modelled as $Q_t = Q^{_F}_{t} + Q^{_S}_{t}$ and can be transformed to a volume per unit time (consistent with observations of flow) by multiplying by the catchment area.

In its entirety, the RRM has eight parameters, but following the approach used by \cite{Schoups2010a} we fix $\alpha_S$ to $1\times 10^{-6}$ resulting in a percolation rate that is effectively linearly related to the storage $U_t$.  Fixing this parameter, truncates the parameter vector to $\theta = (I_{\mathrm{max}}, U_{\mathrm{max}}, Q^{_s}_{\mathrm{max}}, \alpha_F, \alpha_E, K_F, K_S)^{\top}$ .

\cite{Schoups2010a} calibrate the model above to daily precipitation, evaporation and streamflow data for the Guadalupe River basin at Spring Branch, Texas, USA.  These authors devised plausible ranges for the seven parameters outlined above, which we provide in Table \ref{tab:paramRangeTable}.

\begin{table}
 \caption{Parameters of the rainfall-runoff model and prior uncertainty ranges.} \label{tab:paramRangeTable}
\begin{center}
  \begin{tabular}{ | l | c  c  c  c | }
    \hline
    Parameter & Symbol & Lower & Upper & Units \\
    \hline
    Maximum interception & $I_{\mathrm{max}}$ & 1 & 10 & mm  \\ 
    Soil water storage capacity & $U_{\mathrm{max}}$ & 10 & 1000 & mm  \\ 
    Maximum percolation rate & $Q^{_s}_{\mathrm{max}}$ & 0 & 100 & mm/day  \\ 
    Evaporation parameter & $\alpha_E$ & 1E-6 & 100 & --  \\ 
    Runoff parameter & $\alpha_F$ & -10 & 10 & --  \\ 
    Time constant, fast reservoir & $K_F$ & 0 & 10 & days  \\ 
    Time constant, slow reservoir & $K_S$ & 0 & 150 & days \\ 
    \hline
  \end{tabular}
\end{center}
\end{table}

To quantify the discrepancy between the data and simulation, denoted $\rho_p(\theta)$, we consider the following relative distance
\begin{align*}
\rho_p(\theta) &= \sum_{t=1}^T \frac{(y_{\mathrm{obs}}^t - y_{\theta}^t)^2}{y_{\mathrm{obs}}^t},
\end{align*}
where $y_{\mathrm{obs}}^t$ and $y_{\theta}^t$ are the observed and simulated streamflow at time $t$.  The number of time points is $T$, where $T=1827$ here.  Matlab code for this hydrology model is available as Example 6 in the DREAM package \citep{Vrugt2016}, which is available at \\ \url{http://www.pc-progress.com/en/onlinehelp/dream1/DREAM_Suite.html?Demoexamples.html}. 

For the history matching procedure, the implausibility for each sample is computed as $\mathcal{I}(\theta) =  \rho_p(\theta) - r \times s_d(\theta)$ where $\rho_p(\theta)$ and $s_p(\theta)$ is the mean prediction of the function and the standard deviation from the currently fitted GP, respectively.  The exploration parameter is set at $r=3$.  The cut-off for implausibility at wave $w$ is given by $\mathcal{I}_w(\theta) > c_w$ where $c_w$ is selected adaptively such that exactly half of the surviving samples satisfy the cut-off.

Even for moderate dimensional problems the brute force approach to history matching is infeasible.  Thus we explore our SMC approach for this task.  We use $M=10$K particles in the SMC and $N=1$K training samples for the GP at each wave.  For illustrative purposes we use 29 waves.

\begin{figure}[!htp]
	\centering
	\includegraphics[height=0.4\textheight,width=0.7\textwidth]{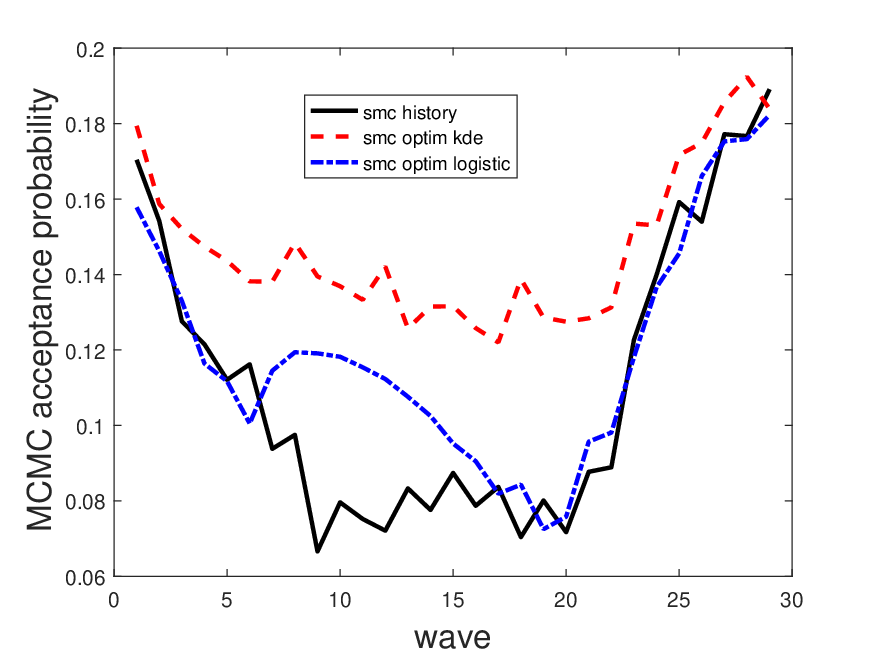}
	\caption{Acceptance probability of the MCMC step when applying the SMC history matching method (solid), SMC optimisation method with kde transforms for the marginals (dash) and the SMC optimisation method with logistic transforms for the marginals (dot-dash).}
	\label{fig:hydrology_smc_acceptance_rates}
\end{figure}

\begin{figure}[!htp]
	\centering
	\includegraphics[height=0.7\textheight,width=0.9\textwidth]{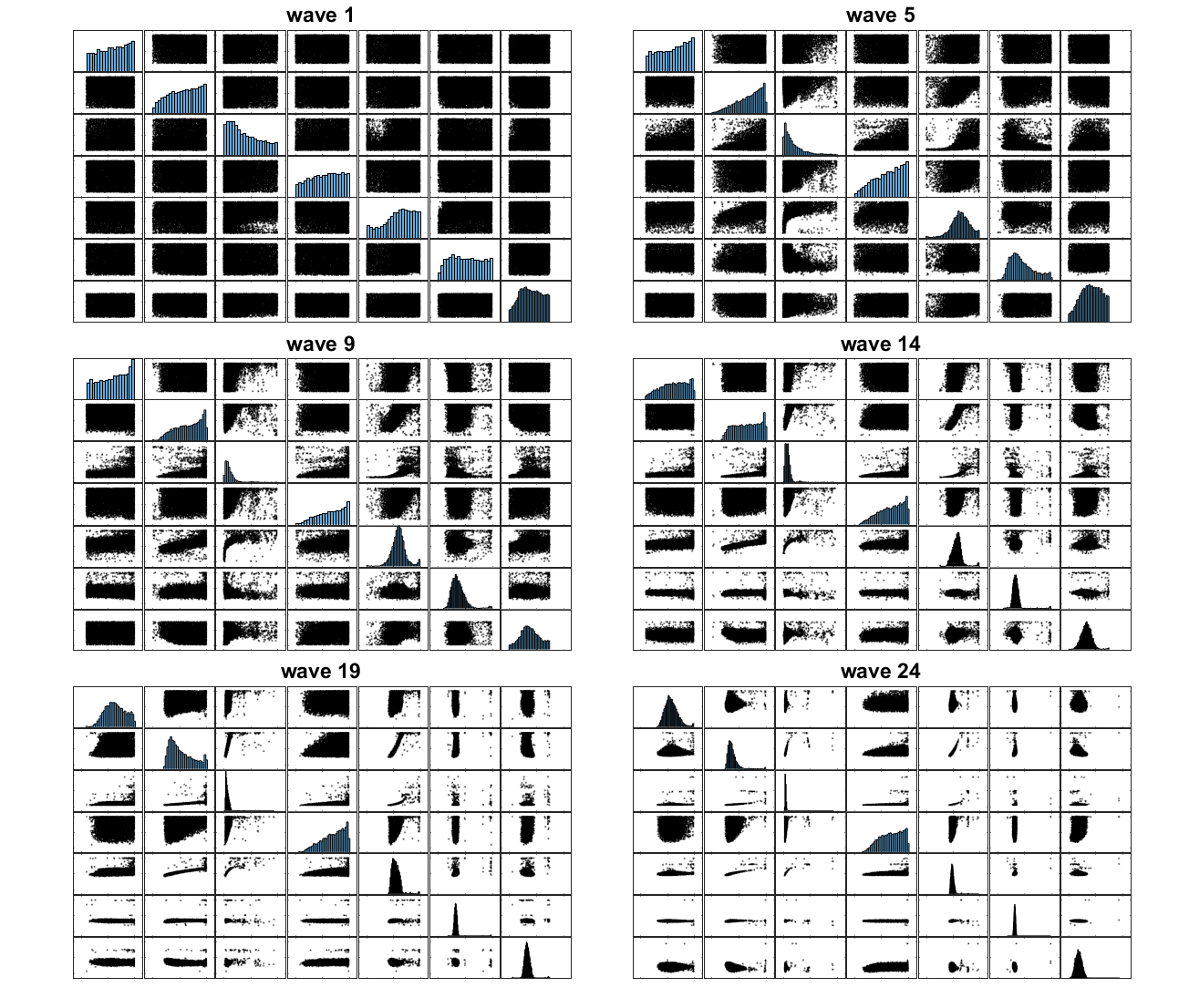}
	\caption{Bivariate scatterplots of the parameters (with marginal histograms along the diagonals) when the SMC history matching method is applied to the hydrology example. }
	\label{fig:hydrology_smc_history_match_matrixplot}
\end{figure}

The SMC procedure appears to be successful in performing the history matching.  However, the acceptance rates of the MCMC step are around 7-18\%, which are significantly smaller than that for the toy example above.  The evolution of the MCMC acceptance probability over the waves is shown as the solid plot in Figure \ref{fig:hydrology_smc_acceptance_rates}.  The acceptance rate of 7\% might be considered small for a seven dimensional parameter space, which indicates that the non-implausible region is difficult to explore effectively.  This is confirmed in Figure \ref{fig:hydrology_smc_history_match_matrixplot}, which shows bivariate plots of the SMC samples after various waves.  It is evident that even the marginal and bivariate distributions are complex, with samples also appearing in disconnected  regions.  In particular, some of the boundaries of the parameter space cannot be ruled out as implausible early in the process.

It turns out that a careful design of the MCMC proposal is required for the SMC procedure to perform well enough.  We tried also using the logistic function and the cdf of the beta distribution (after scaling) to transform the marginals.  However, given the significant multi-modality present, we found that these approaches resulted in a very small MCMC acceptance probability that decreased rapidly.  The value of $R$ became too large for the SMC sampler to be computationally feasible.  This further demonstrates the significantly complex sampling problem that arises from history matching.

\begin{figure}[!htp]
	\centering
	\includegraphics[height=0.7\textheight,width=0.9\textwidth]{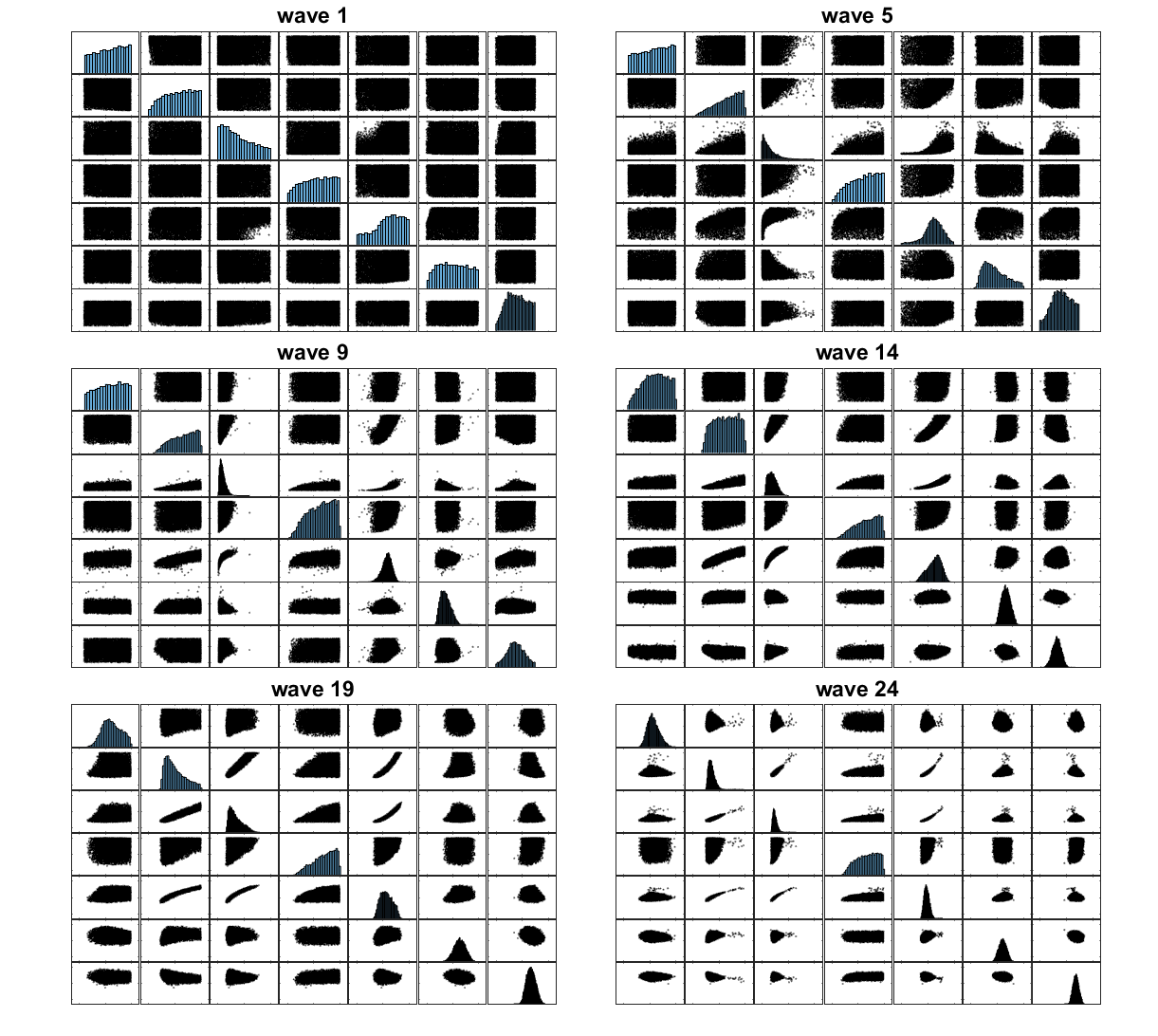}
	\caption{Bivariate scatterplots of the parameters (with marginal histograms along the diagonals) when  SMC optimisation is applied to the hydrology example. }
	\label{fig:hydrology_smc_optim_logistic_matrixplot}
\end{figure}

For comparison purposes we also run a standard SMC optimisation approach (i.e.\ no emulator) with $M=10$K particles.  The sequence of distributions is defined by the 0.5 quantile of the distances in the current particle set.  The same type of MCMC kernel is used in the move step.  We again perform 29 waves.  The bivariate plots after certain iterations of this approach are shown in Figure \ref{fig:hydrology_smc_optim_logistic_matrixplot}.  It is evident, when comparing to Figure \ref{fig:hydrology_smc_history_match_matrixplot}, that relatively regular bivariate distributions of the parameters arise from the SMC optimisation.  Hence, the complexity of the joint distributions as seen in Figure \ref{fig:hydrology_smc_history_match_matrixplot} is mostly an artifact of the history matching method rather than being a result of the hydrology model.  The difficulty in sampling the parameter space is in part reflected in the acceptance probability of the MCMC kernel of the SMC history matching and optimisation approaches.  Using the same type of proposal in the MCMC kernel, it is clear from Figure \ref{fig:hydrology_smc_acceptance_rates} that SMC optimisation has a much higher acceptance rate than history matching (compare the solid plot with the dash plot).  Furthermore, the distributions arising from SMC optimisation are regular enough that it is not necessary to resort to using kde fitting of the marginals as presented in Algorithm \ref{alg:move}.  Figure \ref{fig:hydrology_smc_acceptance_rates} shows that a simple logistic transform of the marginals leads to reasonable acceptance rates (dot-dash plot).

Of course these results do not imply that history matching should be dismissed; over the 29 waves, history matching uses more than two orders of magnitude fewer simulations of the model, which is critical in applications where history matching is adopted.  However, the results do highlight the challenges associated with history matching and how SMC history matching can help to address those challenges.

The above hydrology example uses a deterministic model, whereas the next example considers a stochastic process and the implausibility measure depends on an unbiased likelihood estimator.

\subsection{Gene Network Example}

\citet{Golightly2005}, and more recently \citet{GPPseudoMarginal2017},  consider a Markov jump process for an autoregulatory gene network consisting of four species DNA, RNA, P and P$_2$.  See \citet{Golightly2005} for more details.  The system contains eight possible reactions
\begin{align*}
\mathrm{DNA} + \mathrm{P}_2 & \xrightarrow{c_1 \mathrm{DNA} \times \mathrm{P}_2} \mathrm{DNA}\cdot \mathrm{P}_2,& 2\mathrm{P} & \xrightarrow{c_5 \mathrm{P}(\mathrm{P}-1)/2}  \mathrm{P}_2, \\
\mathrm{DNA}\cdot \mathrm{P}_2 & \xrightarrow{c_2 (k-\mathrm{DNA})} \mathrm{DNA} + \mathrm{P}_2, & \mathrm{P}_2 & \xrightarrow{c_6 \mathrm{P}_2}  2\mathrm{P},\\
\mathrm{DNA} & \xrightarrow{c_3 \mathrm{DNA}} \mathrm{DNA} + \mathrm{RNA}, & \mathrm{RNA} & \xrightarrow{c_7 \mathrm{RNA}}  \emptyset, \\
\mathrm{RNA} & \xrightarrow{c_4 \mathrm{RNA}}\mathrm{RNA} + \mathrm{P}, & \mathrm{P} & \xrightarrow{c_8 \mathrm{P}}  \emptyset,
\end{align*}
where $k$ is a conservation constant (number of copies of the gene)
and $\vect{c}=(c_1,\ldots,c_8)$ are the rate constants
governing the speed at which the system evolves. We consider the same scenario
in \citet{Golightly2005} where data are simulated using rate values
$\vect{c}=(0.1,0.7,0.35,0.2,0.1,0.9,0.3,0.1)$, with $k=10$, and
initial species levels
$(\mathrm{DNA},\mathrm{RNA},\mathrm{P},\mathrm{P}_2)=(5,8,8,8)$. We
simulate equi-spaced data as the next 100 observations (on all
species) recorded at 0.5 unit time intervals.   Note that we assume these data are observed without error. As in \citet{Fearnhead2014}, we take independent half-Cauchy priors for the parameters, with density $p(c_i)\propto 1/(1 + 4c_i^2)$, $c_i>0$ for $i=1,\ldots,8$. We remove the positivity constraint on the rate parameters by working on the log scale, that is, with $\theta_i=\log c_i$ for $i=1,\ldots,8$.

This model does not have a computationally tractable likelihood function.  One approach to perform inference for such models is to assume that each species is observed with Gaussian error with a standard deviation of $\sigma$ \citep{Holenstein2009}.  The likelihood of the implied state space model can be estimated with a particle filter using $J$ particles \citep{Gordon1993}.  More accurate inferences are obtained with a low value of $\sigma$, with the correct posterior obtained in the limit as $\sigma\to 0$. However as $\sigma$ decreases, more particles $(J)$ are required to obtain an accurate likelihood estimate, increasing the computation.  For illustration purposes, we use $\sigma = 0.6$ and $J=6000$ here.

For the history matching procedure, the emulator is trained on values of $\log (-\log \hat{f}(y_{\mathrm{obs}}|\theta))$ as the output, where $\hat{f}(y_{\mathrm{obs}}|\theta)$ is the estimated likelihood.  We take the log twice to improve the capacity of the emulator to provide a good fit (as advocated by \citet{Wilkinson2014}).  Since $\pi(\theta)$ is relatively vague, many datasets from the initial predictive distribution are far from the observed data.  When some rate parameters are relatively large, simulation from the model can be very expensive.  To overcome this, if any of the four species reaches a population size of 100 or the simulation is taking too long, the likelihood estimation procedure is terminated early and $\log \hat{f}(y_{\mathrm{obs}}|\theta)$ is set to the smallest properly estimated log-likelihood initially drawn from $\pi(\theta)$, which is relatively very small.  This discourages exploring such areas of the parameter space in future waves.  For parameter configurations that lead to an estimated likelihood of numerically 0, we also set the log-likelihood estimate to the small value mentioned above.   The implausibility for each sample is computed as $\mathcal{I}(\theta) =  y_p(\theta) - r \times s_d(\theta)$ where $y_p(\theta)$ and $s_p(\theta)$ is the mean prediction and the standard deviation of the predicted output from the currently fitted GP, respectively.  The exploration parameter is set at $r=3$.  The cut-off for implausibility at wave $w$ is given by $\mathcal{I}_w(\theta) > c_w$ where $c_w$ is selected adaptively such that exactly half of the surviving samples satisfy the cut-off.

\begin{figure}[!htp]
	\centering
	\includegraphics[height=0.5\textheight,width=1.0\textwidth]{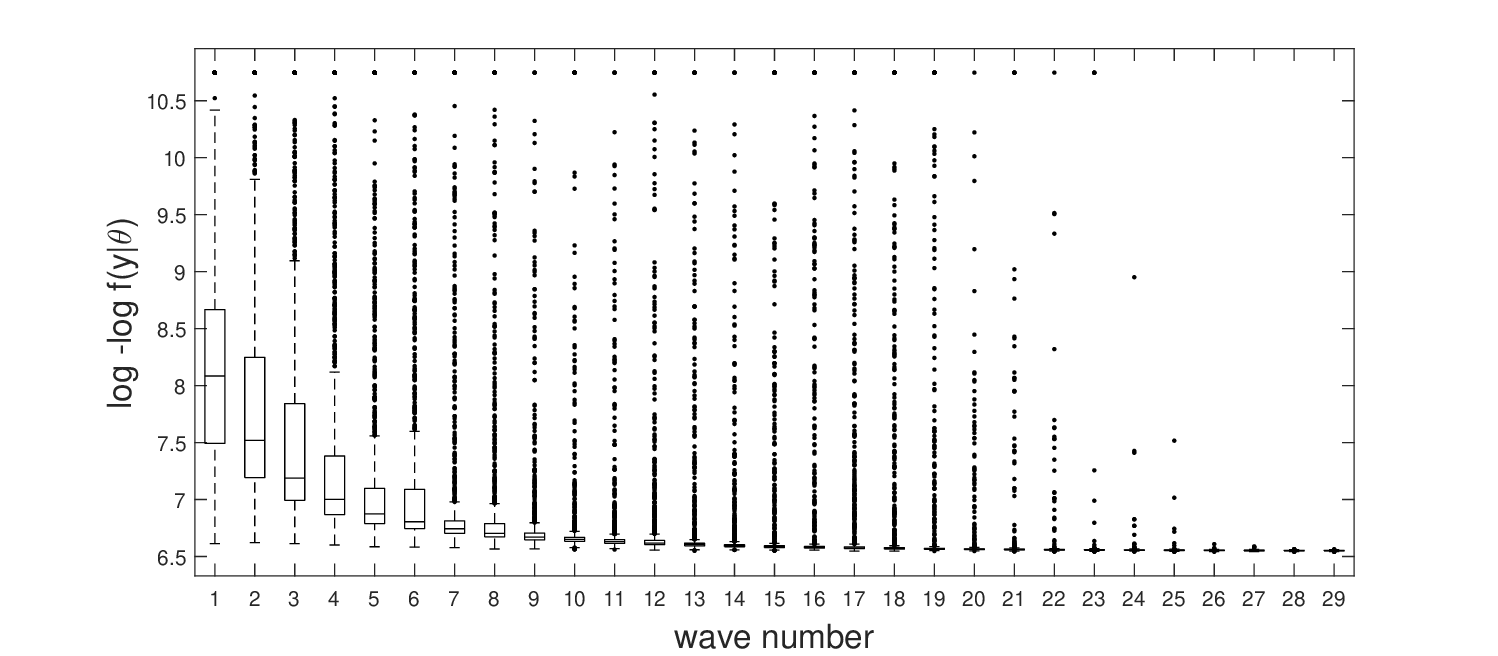}
	\caption{Boxplots of the output $\log (-\log \hat{f}(y_{\mathrm{obs}}|\theta))$ from the training points at each wave for the gene network example.  The largest value of the outliers typically corresponds to a parameter value that simulates data far from the observed data and/or can take an excessive time to simulate.}
	\label{fig:gene_boxplot_output}
\end{figure}

We perform 30 waves of our SMC history matching method.  Boxplots of the output $\log (-\log \hat{f}(y_{\mathrm{obs}}|\theta))$ from the training points at each wave are shown in Figure \ref{fig:gene_boxplot_output}.  Even after 20 waves, some of the training points generated by the SMC history matching method are not consistent with the data.

\begin{figure}[!htp]
	\centering
	\includegraphics[height=0.5\textheight,width=1.0\textwidth]{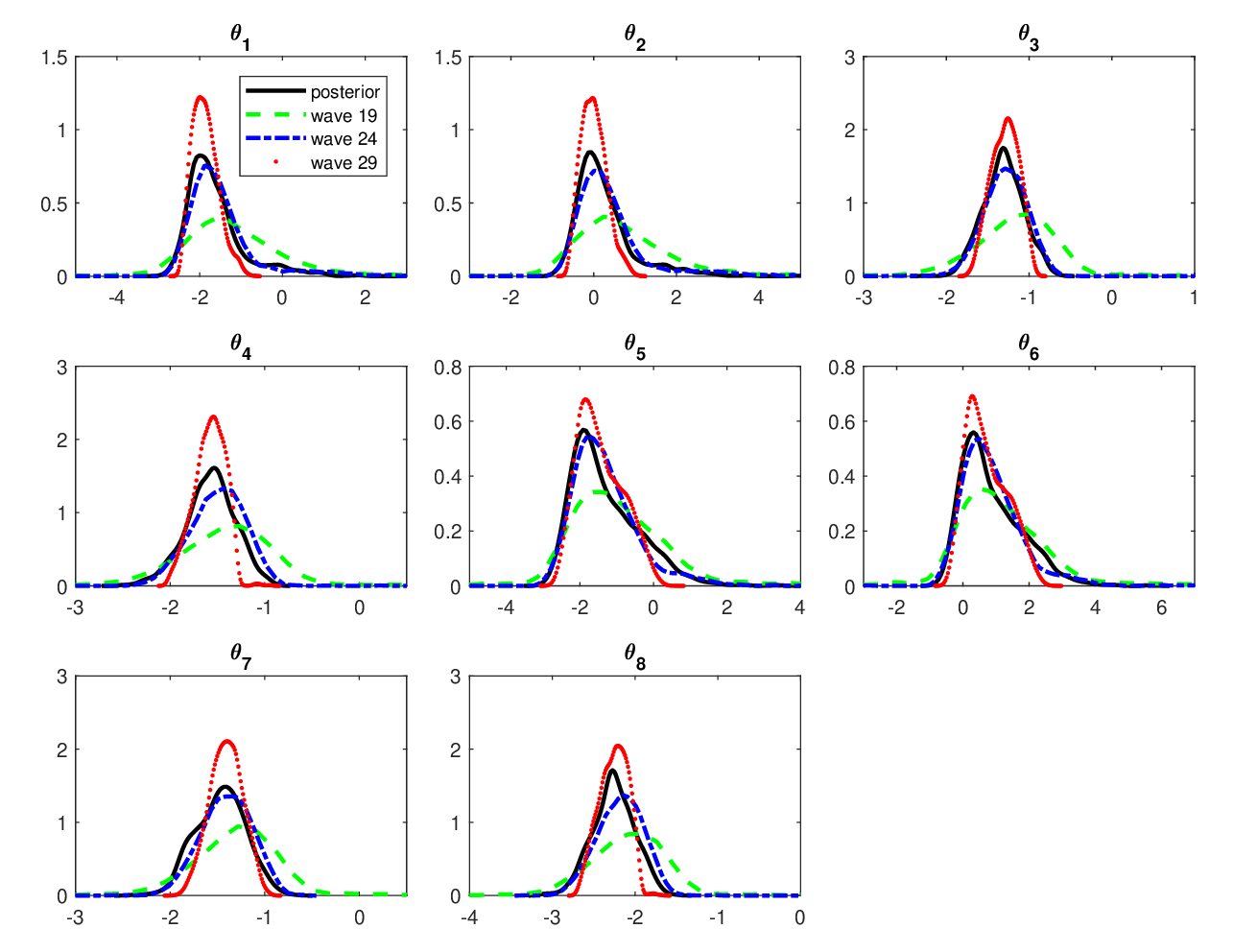}
	\caption{Marginal distributions obtained for the gene network example.  Shown are the marginal posterior distributions from Bayesian SMC (solid), and the marginal distributions obtained from the output of SMC history matching after 19 (dash), 24 (dot-dash) and 29 (dot) waves. }
	\label{fig:gene_marginal_distributions}
\end{figure}

\begin{figure}[!htp]
	\centering
	\includegraphics[height=0.4\textheight,width=0.7\textwidth]{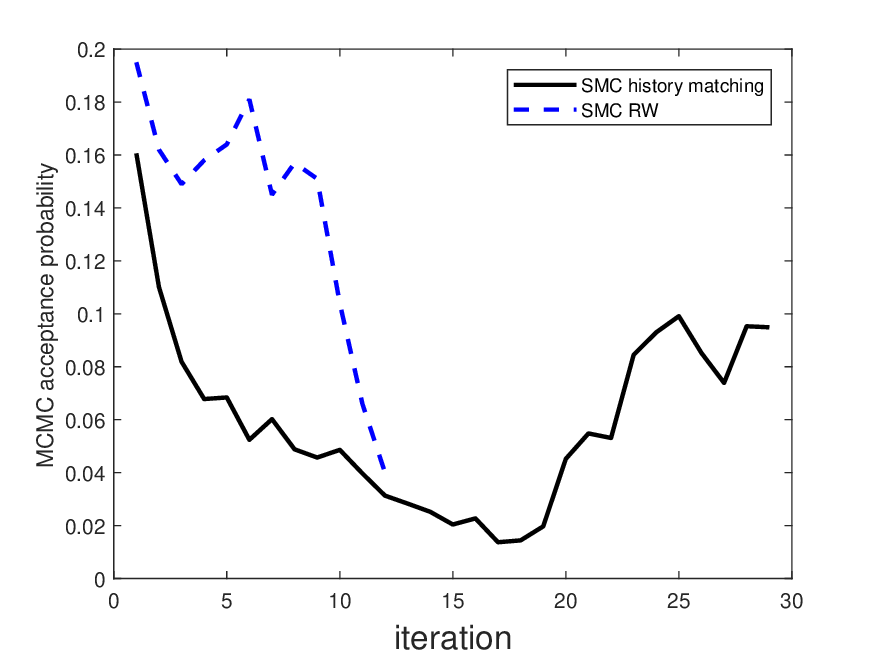}
	\caption{Acceptance probability of the MCMC step when applying the SMC history matching method (solid) and the Bayesian SMC method with a multivariate normal random walk (RW) proposal (dash) to the gene network example.}
	\label{fig:gene_smc_acceptance_rates}
\end{figure}

For comparison purposes we also run a standard SMC sampler with 1K particles to sample from the posterior distribution.  The SMC sampler we use is based on likelihood annealing and uses the adaptive features described in \citet{South2016}, which adaptively determines the annealing temperatures and the number of MCMC repeats in the move step as we find it works well enough here.  Here we use a multivariate normal random walk proposal in the move step.  The marginal posterior distributions, together with the marginal distributions obtained from SMC history matching after 19, 24 and 29 waves, are shown in Figure \ref{fig:gene_marginal_distributions}.  The SMC procedure only requires 11 intermediate temperatures to reach the posterior, which are roughly $0.0004$, $0.0014$, $0.005$, $0.0127$, $0.028$, $0.056$, $0.104$, $0.18$, $0.30$, $0.48$ and $0.74$.  In contrast, even after 24 waves the distributions implied by SMC history matching are less precise than the posterior.  This highlights that a relatively large number of waves are required to eliminate poor parts of the parameter space.  Figure \ref{fig:gene_smc_acceptance_rates} plots the MCMC acceptance rate of the history matching and Bayesian SMC methods over the iterations.  It is clear that the Bayesian SMC method produces a higher acceptance rate, even with a simpler proposal distribution.  This again highlights the difficult sampling problem generated by history matching.

Despite the above comparisons, history matching only uses 30K model simulations (stopping after wave 24, which might be considered reasonable, would result in 25K model simulations) whilst the Bayesian SMC method uses roughly 460K model simulations with only 1K SMC particles (SMC history matching had 10K particles). 

\subsection{Reservoir Modelling Example}

The previous examples consider only a single output, which is either a distance or likelihood.  However, history matching is often applied in situations where there is interest in matching on  multiple outputs, or a set of summary statistics of the data.  For the purposes of demonstrating the latter, we consider the IC Fault model (e.g.\ \citet{Tavassoli2005}), which was considered as a benchmark example for history matching in \citet{Salter2016}.

The IC fault model is a cross-sectional model of a reservoir.  The model contains three unknown parameters; $h$, (the fault throw), $k_g$ (the good-quality sand permeability), and $k_h$ (the poor-quality sand permeability).  The model produces a multivariate time series of length 36 months, consisting of the oil production rate, water injection rate and water cut (or production) rate recorded each month.  Following \citet{Salter2016}, we attempt to match on three statistics: $o_{24}$ (the oil production rate in Month24), $o_{36}$ (the oil production rate in Month 36)  and $w_{36}$ (the water injection rate in Month 36).  The observed value of the statistic is assumed to be $(563.6, 387.5, 917.2)^\top$. 

The model has previously been run at 159,661 different parameter values in the space $h \in (0,60)$, $k_g \in (100,200)$ and $k_h \in (0,50)$.  The parameter values tested and the corresponding model output can be downloaded from \\ \url{http://www.imperial.ac.uk/earth-science/research/research-groups/perm/standard-models/}.  For simplicity, instead of simulating directly from the model at an untried parameter value, we assume that the model output can be well approximated by taking the model output at the closest parameter value in Euclidean space within the pre-existing database.  We take this approach to produce the ``expensive" model runs.  The model output is thus inherently discrete as it is only recorded to one decimal place and there are many repeated values.

In the history matching, the three model summary statistics are emulated with a separate GP.   The implausibility measure for each output is given by \eqref{eq:implausibility}, where only emulation uncertainty is considered.  We take the overall implausibility measure as the maximum of the three implausibility values.  We use $M=10,000$ particles in the SMC and $N=100$ training samples for the GP at each wave.  As per the previous examples, we compare SMC history matching with an SMC optimisation approach that uses $10,000$ particles.  Here the distance to be minimised is taken as the Euclidean distance between the model output and observed summaries.  We terminated the SMC history matching when the acceptance rate for the MCMC step drops below 1\%, which occurs after at the 18th wave.  We also run SMC optimisation for 18 waves.   After the 18th wave, SMC optimisation has around 40\% of its particles with the lowest Euclidean distance.

\begin{figure}[!htp]
	\centering
	\includegraphics[height=0.7\textheight,width=0.9\textwidth]{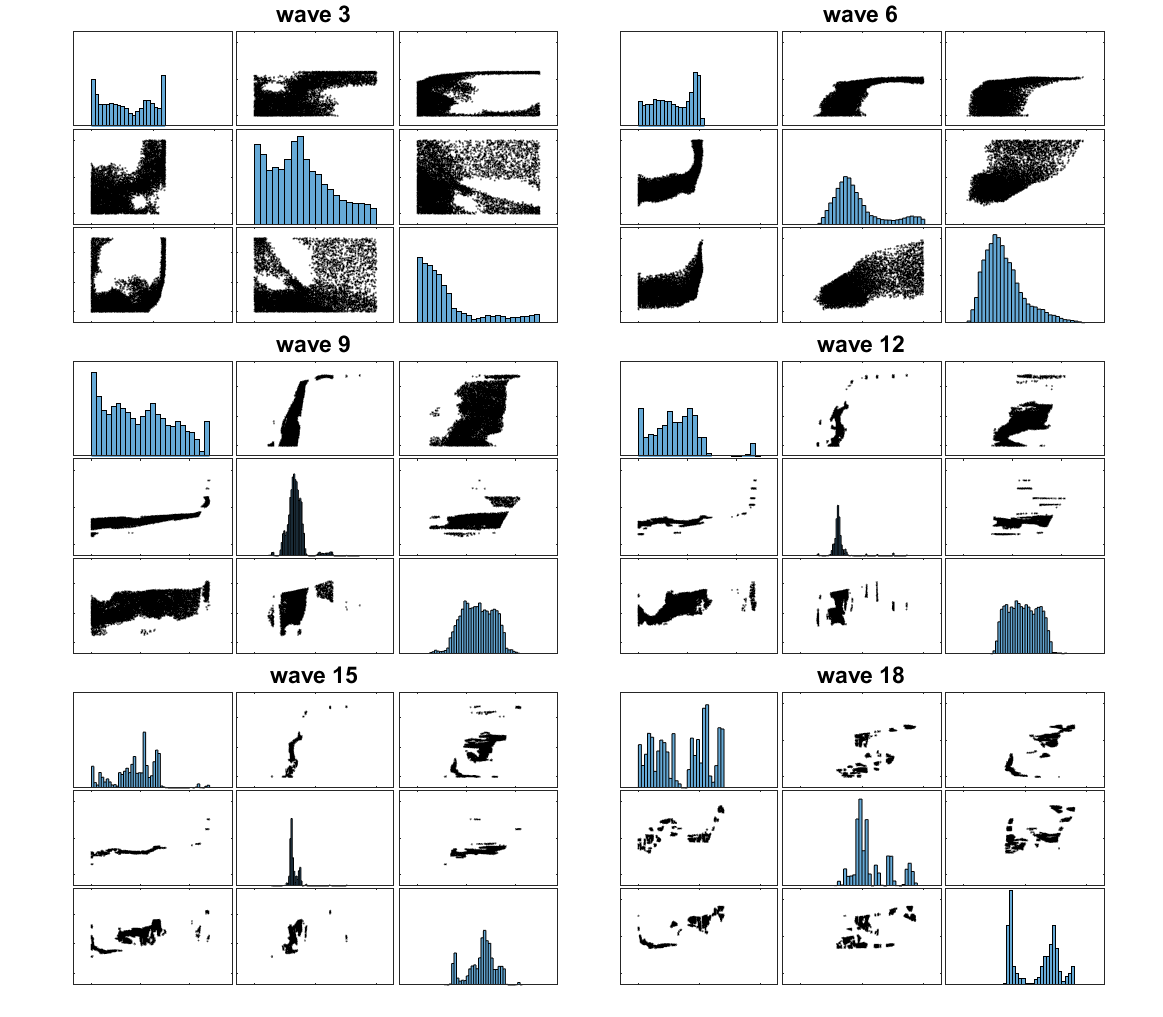}
	\caption{Bivariate scatterplots of the parameters (with marginal histograms along the diagonals) when  SMC history matching is applied to the IC Fault example. }
	\label{fig:IC_fault_history}
\end{figure}

\begin{figure}[!htp]
	\centering
	\includegraphics[height=0.7\textheight,width=0.9\textwidth]{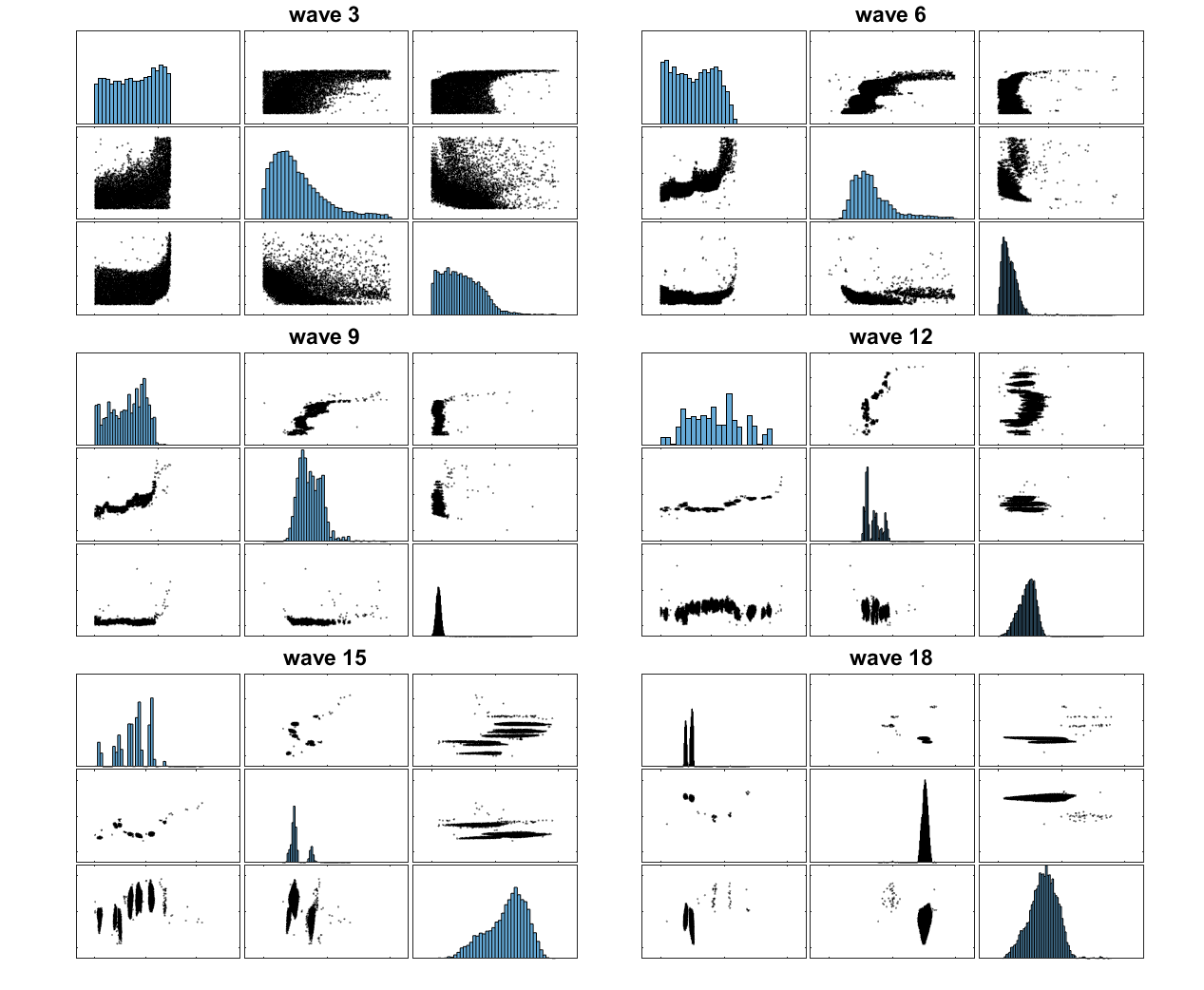}
	\caption{Bivariate scatterplots of the parameters (with marginal histograms along the diagonals) when  SMC optimisation is applied to the IC Fault example. }
	\label{fig:IC_fault_optim}
\end{figure}

Bivariate scatterplots of the particles from SMC history matching and optimisation for various waves are shown in Figures \ref{fig:IC_fault_history} and \ref{fig:IC_fault_optim}, respectively.  The SMC optimisation approach demonstrates that there are different pockets of the parameter value that lead to close matches with the observed statistic.  However, the history matching method appears to generate non-implausible regions with additional multimodality.  Even our carefully designed and sophisticated MCMC kernel described in Section \ref{sec:SMC_history_match}  is having great difficulty sampling from the non-implausible space, further highlighting the challenging sampling problem that history matching can generate.

%\begin{figure}[!htp]
%	\centering
%	\includegraphics[height=0.3\textheight,width=0.4\textwidth]{IC_fault_accrate}
%	\caption{Acceptance probability of the MCMC step when applying the SMC optimisation (solid) and SMC history matching (dash) methods to the IC Fault example.}
%	\label{fig:IC_fault_accrate}
%\end{figure}

\begin{figure}[!htp]
	\centering
	\includegraphics[height=0.3\textheight,width=1.0\textwidth]{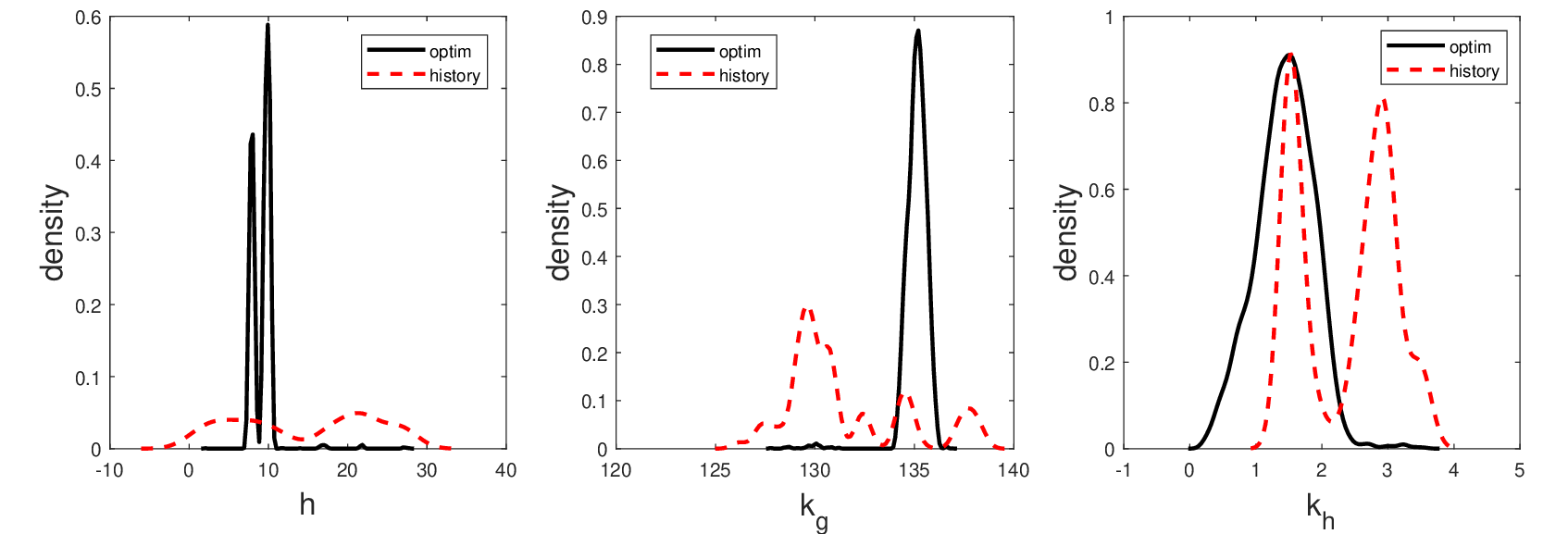}
	\caption{Marginal distributions estimated from the samples obtained with SMC optimisation (solid) and SMC history matching (dash) at wave 18 for the IC Fault example.}
	\label{fig:IC_fault_parameters}
\end{figure}

\begin{figure}[!htp]
	\centering
	\includegraphics[height=0.3\textheight,width=0.4\textwidth]{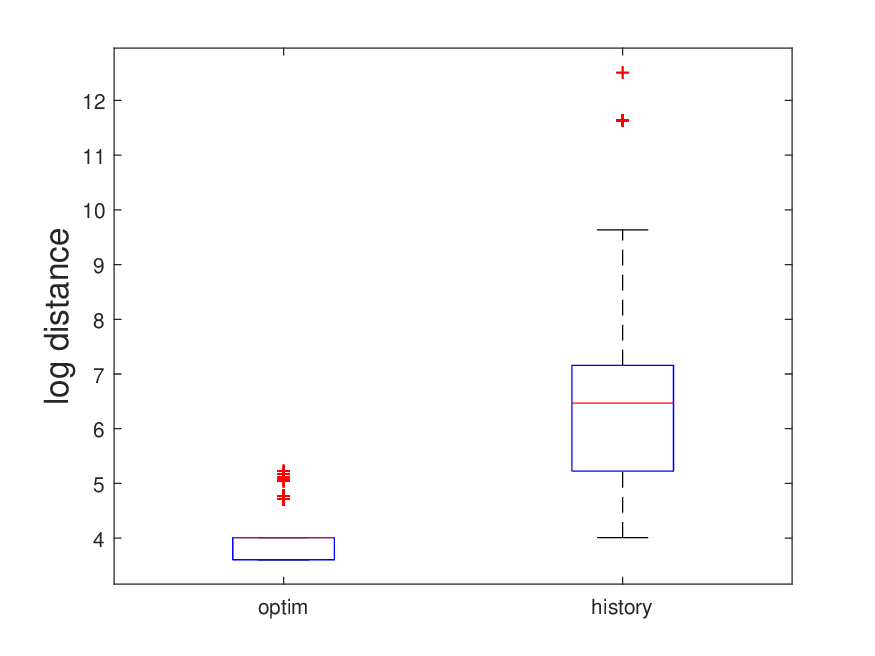}
	\caption{Distances from the model outputs and observed summaries based on the samples obtained with SMC optimisation and SMC history matching at wave 18 for the IC Fault example.}
	\label{fig:IC_fault_distance}
\end{figure}

Figures \ref{fig:IC_fault_parameters} and \ref{fig:IC_fault_distance} demonstrate that, unsurprisingly given the additional expensive models runs, that SMC optimisation finds a more narrow volume of the parameter space that generate close matches with the observed outputs at wave 18.

\section{Discussion} \label{sec:Discussion} 

In this paper we have developed a novel algorithm based on SMC for history matching that is adaptive and offers a principled method for sampling from the non-implausible space at each wave.  Our algorithm reveals in greater detail the significant complexity of the probability distribution associated with the non-implausible space that arises from history matching.  

We advise practitioners that more care and computation may be required to help ensure that potentially important pockets of the parameter space are not ignored.  Clearly history matching is an important method in the context of expensive simulators.  \citet{McKinley2017} demonstrate in a complex stochastic epidemic model that history matching is able to determine parameter regions that lead to close matches with observed outputs whereas a more conventional SMC ABC algorithm that requires significantly more model simulations is not able to in a feasible amount of time.  More sophisticated emulation approaches than that used in this paper, such as including a mean structure and fitting separate emulators for each disconnected region (see \citet{Vernon2010}), can help to more quickly discard implausible pockets of the parameter space, reducing the multimodality problem.  However, even with more accurate emulation modelling, it is likely that the non-implausible region remains difficult to sample generally.

Our method is semi-automatic in that it adaptively chooses the implausibility thresholds and the number of MCMC repeats, but the practitioner is still required to choose a suitable proposal distribution for the MCMC kernel.  Even in our applications of moderate dimension we found that a simple multivariate normal random walk, which is commonly used in the SMC literature (e.g.\ \citet{Chopin2002} and \citet{DrovandiPettittSCID2011}), is not efficient enough.  Here we developed a method based on transforming the marginals with kde estimates before applying a multivariate normal random walk.  Although we found some success with this approach, the MCMC acceptance rates were sometimes small, which significantly increased computational burden.  An extension of the slice sampling method used in \citet{Andrianakis2017} may prove useful.  The SMC approach of \citet{Schuster2017}, called kernel SMC, uses estimated derivatives and local covariance estimates to improve sampler efficiency.  In summary,  more research in determining a suitable MCMC kernel that is able to explore irregular, multi-modal distributions without derivative information, which can arise from history matching, is required.

In this paper, the emulator for wave $w$ is trained only on points sampled from the non-implausible space at that wave.  However, it is possible to apply the emulator to all training points generated up to wave $w$.  An alternative approach to training emulators is that rather than proceeding in waves of distinct training sets, training samples are sequentially appended to an existing training set stemming from a method called Bayesian optimisation \citep{Mockus2012}.    We might be interested in placing additional training points where there is significant uncertainty in the emulator's prediction.  It is important to note though, that with certain emulators such as GPs the prediction cost is $\mathcal{O}(N)$ where $N$ is number of training points.  See \citet{Holden2015} for further discussion on different emulator training strategies.  We suggest that SMC might also be useful in this sequential training approach, and we leave that for future research.  

To simplify the explanation of the ideas of this paper we assumed that only one output required emulation.  However, in many realistic applications (e.g.\ \citet{Vernon2014} and \citet{Andrianakis2015}) there are multiple outputs which are emulated separately.   One possible approach as described in \citet{Andrianakis2015} is to take the overall implausibility measure as the maximum implausibility over all outputs.  In this case an appropriate implausibility cut-off threshold is even less clear, which further motivates our semi-automated method.  Another approach is to introduce outputs in some sequential fashion, focussing initially on the outputs that are emulated accurately.  It is possible to extend our SMC approach to accommodate this heuristic by modifying the sequence of distributions that SMC depends on to include an indicator highlighting which outputs have been introduced.  One approach would sequentially introduce outputs one-at-a-time once the already introduced outputs are satisfied.  The output to introduce could be adaptively chosen by selecting the one that is most accurately emulated out of the outputs not yet introduced.  The implausibility thresholds for each introduced output can be determined adaptively using the same approach as in our paper.

\section*{Acknowledgements}

 Christopher Drovandi was supported by an Australian Research Council Discovery Project (DP200102101).   David Nott was supported by a Singapore Ministry of Education Academic Research Fund Tier 2 grant (MOE2016-T2-2-135).  Christopher Drovandi is grateful to Leah South for many discussions on sequential Monte Carlo methods and for providing some useful code.  Christopher Drovandi is an Associate Investigator of the Australian Centre of Excellence for Mathematical and Statistical Frontiers.   David Nott is affiliated with the Operations Research and Analytics Research cluster, National University of Singapore.

\bibliographystyle{apalike} 
\bibliography{refs}

\end{document}